\documentclass[12pt]{article}
\pdfoutput=1

\newlength{\abstractwidth}
\usepackage{epsf}
\usepackage{color}
\usepackage{graphicx}

\usepackage{amssymb}
\usepackage{latexsym}

\renewcommand{\thefootnote}{\fnsymbol{footnote}}
\renewcommand{\thanks}[1]{\footnote{#1}}
\newcommand{\starttext}{
\setcounter{footnote}{0}
\renewcommand{\thefootnote}{\arabic{footnote}}}

\newcommand{\bea}{\begin{eqnarray}}
\newcommand{\eea}{\end{eqnarray}}
\newcommand{\be}{\begin{eqnarray}}
\newcommand{\ee}{\end{eqnarray}}
\newcommand{\<}{\langle}
\renewcommand{\>}{\rangle}

\def\a{\alpha}
\def\b{\beta}
\def\g{\gamma}

\def\tet{\vartheta}
\def\ep{\varepsilon}

\def\mN{\mathfrak{N}}

\def\cD{{\cal D}}
\def\cE{{\cal E}}
\def\cF{{\cal F}}
\def\cG{{\cal G}}

\def\cI{{\cal I}}
\def\cJ{{\cal J}}

\def\cM{{\cal M}}
\def\cN{{\cal N}}
\def\cO{{\cal O}}

\def\cR{{\cal R}}

\def\bZ{{\bf Z}}

\def\bw{{\bf w}}

\def\bz{{\bf z}}

\def\half{ {1\over 2}}
\def\p{\partial}

\def\pbz{\p_{\bar z}}
\def\pbw{\p_{\bar w}}
\def\Sr{{\Sigma_{\rm red}}}
\def\om{\omega}

\def\no{\nonumber}
\def\sm{\smallskip}

\def\mD{\mathfrak{D}}

\def\mM{\mathfrak{M}}
\def\mN{\mathfrak{N}}

\def\mg{\mathfrak{g}}
\def\mh{\mathfrak{h}}

\def\ZZ{{\mathbb Z}}

\def\CC{{\mathbb C}}

\def\half{{1\over 2}}

\def\p{\partial}

\def\gg{\mg}
\def\rh{\hat \rho}

\def\pz{\p_z}
\def\pw{\p_w}

\begin{document}
\starttext
\baselineskip=16pt
\setcounter{footnote}{0}

\begin{flushright}
2015 January 11
\end{flushright}

\bigskip

\begin{center}
{\Large\bf The Super Period Matrix with Ramond Punctures }

\vskip 0.15in 

{\Large \bf  in the supergravity formulation}
\footnote{Research supported in part by National Science
Foundation grants PHY-13-13986 and DMS-12-66033.}

\bigskip\bigskip
{\large Eric D'Hoker$^a$ and Duong H. Phong$^b$}

\bigskip
$^a$ \sl Department of Physics and Astronomy \\
\sl University of California, Los Angeles, CA 90095, USA\\
$^b$ \sl Department of Mathematics\\
\sl Columbia University, New York, NY 10027, USA

\end{center}

\bigskip\bigskip

\begin{abstract}

In a very recent preprint, Witten showed how to construct a $g|r \,  \times \, g|r$ super period matrix for super Riemann surfaces of genus $g$ with $2r$ Ramond punctures, which is symmetric in the ${\bf Z}_2$ graded sense. He also showed how it can be applied to analyze supersymmetry breaking in string compactifications which are supersymmetric at tree-level. Witten's construction is in the purely holomorphic formulation of super Riemann surfaces. In this paper, a construction is given in the formulation of two-dimensional supergravity. The variations of the super period matrix with respect to supermoduli deformations are also given, as well as an explicit illustration of how the super period matrix with two Ramond punctures would emerge from a degeneration of the super period matrix without punctures in higher genus.

\end{abstract}

\vfill\eject


\newpage

\baselineskip=15pt
\setcounter{equation}{0}
\setcounter{footnote}{0}

\section{Introduction}
\setcounter{equation}{0}
\label{sec1}

A problem in superstring perturbation theory which has remained relatively unexplored is scattering amplitudes involving Ramond states. For tree-level and one-loop orders, such amplitudes had been investigated early on in \cite{Friedan:1985ge} and \cite{Atick:1986rs} respectively. But starting from two loops and higher, a new, geometric and fundamental difficulty emerges, which is that of supermoduli. Some of these supermoduli difficulties were overcome at two-loop level for the scattering amplitudes of Neveu-Schwarz states \cite{D'Hoker:2001nj,D'Hoker:2001qp, D'Hoker:2005jc}, but not for Ramond states. Besides its intrinsic interest, a perturbative understanding of scattering amplitudes with Ramond states is also crucial for many other fundamental issues, such as Ward identities for supersymmetry and the perturbative breaking of supersymmetry in theories which are supersymmetric at tree-level. Some of these issues had been successfully addressed already long ago to one-loop order in \cite{Dine:1987xk,Dine:1987gj,Atick:1987gy}
(see also an early general discussion in \cite{Martinec:1986wa}). The problem at higher loop order, however, was significantly clarified only  recently by Witten in \cite{Witten:2012ga,Witten:2012bh,Witten:2013cia}, where the subject was also thoroughly reviewed.  There, it was shown that the presence of Ramond states leads to a novel geometric situation: unlike Neveu-Schwarz  states, which may be placed at marked points on the super worldsheet, the presence of Ramond states changes even the geometric nature of the super worldsheet itself \cite{Witten:2012ga}. 

\sm

An eventual understanding of the scattering of Ramond states in superstring perturbation theory would require then an understanding of super Riemann surfaces with Ramond punctures and their moduli. This question has been raised  essentially for the first time in \cite{Witten:2012ga,Witten:2012bh,Witten:2013cia}, and  little was known prior to these works: is there an analogue of the super period matrix ? is it symmetric ? can the projection methods of \cite{D'Hoker:2001nj,D'Hoker:2002gw} for gauge-fixing the superstring measure be adapted to the case of genus $2$ with Ramond punctures ? These questions have now been answered affirmatively in the very recent preprint of Witten \cite{new.Witten}. A fundamental tool introduced there is the notion of ``fermionic periods" of superholomorphic forms, which can be used to normalize a basis, and leads to a $g|r \, \times \, g|r$ matrix of periods, which is symmetric in the ${\bf Z}_2$ graded sense. The super period matrix with Ramond punctures was then applied to show the vanishing of the bulk contribution to the vacuum energy for heterotic string compactifications which are supersymmetric at tree level. The word ``bulk" refers here to the contributions from the interior of supermoduli space. Thus the vacuum energy receives only contributions from the boundary. The super period matrix with Ramond punctures has actually a pole along a divisor, so that an essential step in \cite{new.Witten} is to show that this pole gets cancelled by the contributions of fermionic zero modes.

\sm

Witten's approach in \cite{new.Witten} is entirely in the holomorphic formulation of super Riemann surfaces described in detail in \cite{Witten:2012ga}. The goal of the present paper is to present an alternative approach in terms of the component formulation of two-dimensional supergravity. As a consequence, we obtain explicit formulas illustrating the dependence of the super period matrix on the underlying reduced surface, the gravitino field, and the Ramond punctures. The divisor of poles for the super period matrix is also identified explicitly in terms of $\tet$-divisors. Such formulas for super Riemann surfaces with Neveu-Schwarz punctures were instrumental in the derivation of the measure \cite{D'Hoker:2001nj,D'Hoker:2001zp} and scattering amplitudes \cite{D'Hoker:2005jc,D'Hoker:2005ht} for Neveu-Schwarz states. Their counterparts for surfaces with Ramond punctures can be expected to play a similar role in future investigations of Ramond states scattering.

\sm

The paper is organized as follows. In \S 2, we begin by reviewing the supergravity approach to super Riemann surfaces. We explain the turning on of the gravitino field $\chi$ and, for the purpose of subsequent publications on the evaluation of Ramond amplitudes, the simultaneous turning on of a Beltrami differential $\mu$, the result of which is to parametrize the fiber of super Riemann surfaces with Ramond punctures over a given reduced surface $\Sr$ with Ramond punctures. We also develop the basic function theory which is needed in the presence of Ramond punctures, namely explicit formulas for the holomorphic sections  of the bundle ${\cal R}^{-1}$ and construction of an anti-symmetric Szeg\"o kernel. In \S 3, we construct a basis of both even and odd superholomorphic forms of U(1) weight $1/2$ using the function theory developed in \S 2. The key issue in the construction of a super period matrix is the prescription of a canonical normalization that would single out a basis, the periods of which would then define the entries of the super period matrix. This is one of the problems solved by Witten in \cite{new.Witten}, and a brief summary of his prescription is given in \S 3.3. Witten's prescription is then implemented in the supergravity formalism in \S 3.4. The resulting periods are worked out in \S 3.5-7. In \S 3.8, we illustrate, in a basic case for Ramond punctures, the relation between superholomorphic forms in the supergravity formalism and holomorphic forms on the underlying reduced surface. Such a relation played an essential role in the calculations of \cite{D'Hoker:2001zp}. In \S 3.9, we discuss briefly an additional set of periods, already known to Witten \cite{Witten.note} but not discussed in \cite{new.Witten}, which are
the ones obtained by considering the line integrals of the normalized basis of superholomorphic 1/2 forms between pairs of Ramond punctures.  In \S 4, we derive the first order variations of all the periods of the superholomorphic 1/2 forms under a deformation of super complex structures. These formulas are needed in the forthcoming work on the derivation of scattering amplitudes for Ramond states. Finally, in \S 5, we show how the super period matrix with 2 Ramond punctures in genus $g$ arises in the non-separating degeneration limit of the super period matrix of a super Riemann surface in genus $g+1$ without R-punctures. Appendices A and B contain various technical definitions and detailed derivations.

\bigskip

{\sl Acknowledgements:} We would like to express our most sincere gratitude to Edward Witten for calling our attention to the whole circle of problems around Ramond punctures, super period matrices, and supersymmetric Ward identities. This work is in many ways a continuation of our work on ${\bf Z}_2\times{\bf Z}_2$ Calabi-Yau orbifolds, which is another question that he suggested to us. We are also very grateful to him for many very helpful discussions on Ramond punctures,  for useful suggestions on a draft of this paper, and for communicating to us very early on his ideas on the super period matrix with Ramond punctures as well as the drafts of the paper \cite{new.Witten} prior to publication.

\newpage

\section{Supergravity formulation with Ramond punctures}
\setcounter{equation}{0}
\label{sec2}

The supergravity Ramond-Neveu-Schwarz formulation of superstring perturbation theory is based 
on two-dimensional supergravity, which governs the field theory on the worldsheets of the strings.  
The fields consist of the metric $\mg$ and the gravitino field $\chi^\pm$, as well as of the fields 
$x^\mu , \psi _\pm ^\mu$ which describe respectively the  bosonic and fermionic string degrees 
of freedom with $\mu = 1, \cdots, D$ in a $D$-dimensional space-time. The fields $\chi^\pm, \psi _\pm ^\mu$ are worldsheet spinors, and the index $\pm$ labels the two  chiralities of the spinors. On a surface of genus $g$,  the spinor fields $\chi^\pm, \psi _\pm ^\mu$  may carry any one of the $2^{2g}$ inequivalent spin structures.

\sm

In the chiral splitting approach \cite{D'Hoker:1988ta,D'Hoker:1989ai}, the starting point for the construction of amplitudes for closed oriented superstrings is the functional integral over the fields $\mg, \chi^\pm, x^\mu, \psi _\pm ^\mu$  on worldsheets of  genus $g$.   In the critical dimension $D=10$ of space-time, gauge symmetries lead to the familiar ghost fields and reduce the integration over $\mg$ and $\chi$ at any given loop order $g$  to a finite-dimensional integral over even and odd moduli of the surface.  The {\sl Chiral Splitting Theorem} \cite{D'Hoker:1989ai} guarantees that this functional integral, at fixed spin structure, fixed internal loop momenta, and fixed metric $\mg$ and gravitino $\chi^\pm$  is the absolute value squared of a chiral amplitude. The chiral amplitude of positive chirality, for example, depends only on the fields $\chi^+$ and on holomorphic deformations of the metric, which we parametrize by the Beltrami differential $\mu$. Type II amplitudes are obtained by pairing independent chiral amplitudes of opposite chirality, 
while Heterotic string amplitudes are obtained by pairing a chiral amplitude of positive chirality with a suitable bosonic amplitude, in both cases supplemented with a suitable GSO summation over spin structures.

\sm

The chiral splitting theorem produces a set of effective field theory rules for the construction of chiral amplitudes at fixed spin structure and fixed internal loop momenta directly in terms of chiral fields $x_+ ^\mu, \psi _+^\mu$ and ghosts $b,c,\beta, \gamma$ as a function of the chiral data $\chi^+$ and $ \mu$ only, without appealing to the complex conjugate fields.  These effective rules may be reformulated in superspace on the worldsheet, thus promoting the worldsheet into a super Riemann surface. Their moduli space may be viewed as the equivalence classes  $(\chi^+, \mu)$ under the action of diffeomorphisms and local supersymmetry transformations. This effective supergravity formulation is now closely related with the purely holomorphic approach to super Riemann surfaces and their moduli space, as formulated in section 3.5.2 of \cite{Witten:2012ga}, and both approaches will produce the same chiral amplitudes.

\sm

The supergravity approach leads to a convenient construction of chiral superstring amplitudes in which the $\chi$ and $\mu$ fields are paired respectively against the worldsheet supercurrent and stress tensor. It was used at two-loop order in the calculation of the measure \cite{D'Hoker:2001nj,D'Hoker:2001zp,D'Hoker:2001it} and scattering amplitudes \cite{D'Hoker:2005jc} of NS states (see also \cite{D'Hoker:2002gw} for a review). The holomorphic formulation of super Riemann surfaces and their moduli space  was used in \cite{Witten:2013tpa}  to give a completely novel and very elegant derivation of the superstring measure at genus 2.

\subsection{Super Riemann surfaces with Ramond punctures}

In the present paper, we shall be interested in using the supergravity and chiral splitting approach to describe super Riemann surfaces with Ramond punctures (or R-punctures for short). The starting point will be a {\sl reduced Riemann surface} $\Sr$ corresponding to a super Riemann surface $\Sigma$ for which all odd coordinates (on the surface and on super moduli space) have been set to zero. The space of super Riemann surfaces may then be described  in terms of the function theory on the ordinary Riemann surface $\Sr$ by turning on the worldsheet gravitino field $\chi$ along with deformations $\mu$ of the metric. 

\sm

In superstring theory, R-punctures are  distinguishable since they will support fermion vertex operators with distinguishable quantum numbers. An R-puncture may be formulated in an infinite number of possible {\sl pictures} \cite{Friedan:1985ge}. As stressed in \cite{Witten:2012ga}, on a super Riemann surface $\Sigma$ of arbitrary genus, an R-puncture in the -1/2 picture corresponds to a Ramond divisor on $\Sigma$ (namely a submanifold of $\Sigma$ of dimension $0|1$), while an R-puncture in the -3/2 picture corresponds to a specific point in the Ramond divisor. Lower picture numbers $-1/2 - k$ with $ k \geq 2$ are allowed as well, but will not be considered here.  

\sm

On a compact super Riemann surface, the number of R-punctures is an even number $2r$ where  $r$ is a positive integer. Since an R-puncture is associated with a quadratic branch point it is natural to partition the set of all R-punctures  into pairs $(p_\eta, q_\eta)$ for $\eta = 1, \cdots, r$, each pair representing a quadratic branch cut. It is natural and it will be useful to define the following sums of Ramond divisors,
\bea
\cF = \cF_p + \cF_q \hskip 1in \cF_p = \sum _{\eta=1}^r p_\eta \hskip 0.5in  
\cF_q = \sum _{\eta=1}^r q_\eta
\eea
The partitioning of R-punctures into pairs also has a natural origin when surfaces with R-punctures are obtained by non-separating degeneration from surfaces of higher genus, including from surfaces of genus $g+r$ without R-punctures, as will be discussed in \S \ref{sec5} below. 

\sm

Since the reduced surface $\Sigma _{\rm red}$ is the starting point for the construction of a super Riemann surface $\Sigma$ in the supergravity approach, we shall give a description of the deformation fields $\mu$ and $\chi$ and associated symmetries from the point of view of the reduced surface $\Sr$.
The generators $v$ of infinitesimal diffeomorphisms in the presence of a Ramond divisor $\cF$, must leave $\cF$  invariant and  therefore be smooth sections of the line bundle $K^{-1}\otimes {\cal O}(-{\cal F})$ over $\Sr$. Throughout, $K$ will denote the canonical bundle over $\Sr$. Similarly, the generators $\xi$ of infinitesimal local supersymmetries must be sections of a line bundle ${\cal R}$ over $\Sr$ which may be defined by the isomorphism, 
\bea
\cR^2 = K^{-1} \otimes \cO(-\cF)
\eea
along with  a {\sl generalized spin structure} needed to specify the square root uniquely.

\sm

The transformation properties under diffeomorphisms $v$ and local supersymmetries $\xi$ 
of the fields $\mu$ and $\chi$ (henceforth, we shall denote the positive chirality component $\chi^+$ simply by $\chi$)  on $\Sr$ dictate the bundles over $\Sr$ of which $\mu$ and 
$\chi$ should be sections,  
\bea
v \in \cR^2 & \hskip 1in& \mu \in \bar K \otimes \cR^2
\no \\
\xi \in \cR ~ && \chi \in \bar K \otimes {\cal R}
\eea
We recall that the action of an infinitesimal diffeomorphism $v$ on $\mu$ is given by $\delta _v \mu = \p_{\bar z} v$, while the action of a local supersymmetry transformation $\xi$  on $\mu$ and $\chi$ 
is given by, 
\bea
\label{susy} 
\delta _\xi \mu & = & \xi \, \chi
\no \\ 
\delta _\xi \chi & = & - 2 \pbz \xi - 2 \mu \pz \xi +(\p_z \mu) \xi
\eea
In either picture -1/2 or -3/2, the divisor $\cF$ is transformed into itself. The difference between the two pictures is that  the generators of local supersymmetries in the $-1/2$ picture are allowed to take arbitrary values on $\cF$ so that supersymmetry acts by an automorphism on $\cF$, while they are required to vanish on $\cF$ in the $-3/2$ picture, so that the individual points in $\cF$ are left invariant.

\subsection{Function theory on the reduced surface $\Sr$}

Our goal is to construct moduli for super Riemann surfaces $\Sigma$ of genus $g$ with $2r$ R-punctures. For this, we need to develop some basic function theory on the underlying reduced surface $\Sigma_{\rm red}$, which is a Riemann surface of genus $g$ with $2r$ R-punctures  obtained by setting to zero all anti-commuting parameters in $\Sigma$. We refer the reader to Appendix \ref{A} for a brief review of standard function theory on compact surfaces without punctures. We shall concentrate here in the body of the text on the new features that arise in the presence of R-punctures.

\sm

The surface $\Sigma_{\rm red}$ has genus $g$ and $2r$ R-punctures partitioned into pairs $(p_\eta,q_\eta)$ with $\eta = 1, \cdots,  r$. As usual (see Appendix A), we fix a canonical homology basis $(A_I,B_I)$ of $H^1(\Sr, \ZZ)$. The canonical bundle $K$ of $\Sigma_{\rm red}$ has ${\rm dim}\,H^0(\Sigma_{\rm red},K)=g$ holomorphic sections.   A canonical basis $\omega_I$ of these holomorphic sections  is determined by the requirement,
\bea
\label{Anorm}
\oint_{A_I}\omega_J=\delta_{IJ}.
\eea
The $g \times g$ period matrix $\Omega_{IJ}$ of $\Sr$ is defined by,
\bea
\Omega_{IJ}=\oint_{B_I}\omega_J.
\eea
The matrix $\Omega$ is symmetric with positive definite imaginary part. Along with the positions of the 
$2r$ R-punctures $(p_\eta,q_\eta)$, it provides the moduli for the surface $\Sigma_{\rm red}$. The Jacobian is defined by $J(\Sigma _{\rm red}) = \CC^g / (\ZZ^g + \Omega \ZZ^g)$.

\sm

To complete the basic function theory on $\Sr$ in the presence of R-punctures, we shall also need the Abel map and the prime form, which are multiple-valued on the surface $\Sigma _{\rm red}$. In order to define them properly, we cut the surface $\Sigma_{\rm red}$ along the homology cycles $A_I, B_I$ for $I=1, \cdots, g$ and parametrize the surface  by a simply connected domain $\Sigma _{\rm cut}$ (see Fig. 1 for $g=2$). We choose a reference point $z_0$ from which the Abel map may be defined  by the integral $\int ^z _{z_0} \omega _I$ as a single-valued function throughout $\Sigma_{\rm cut}$. The prime form $E(z,w)$ may be similarly defined, as is done explicitly in Appendix A. 

\begin{figure}[h]
\centering
\begin{tabular}{c}
\includegraphics[width=120mm]{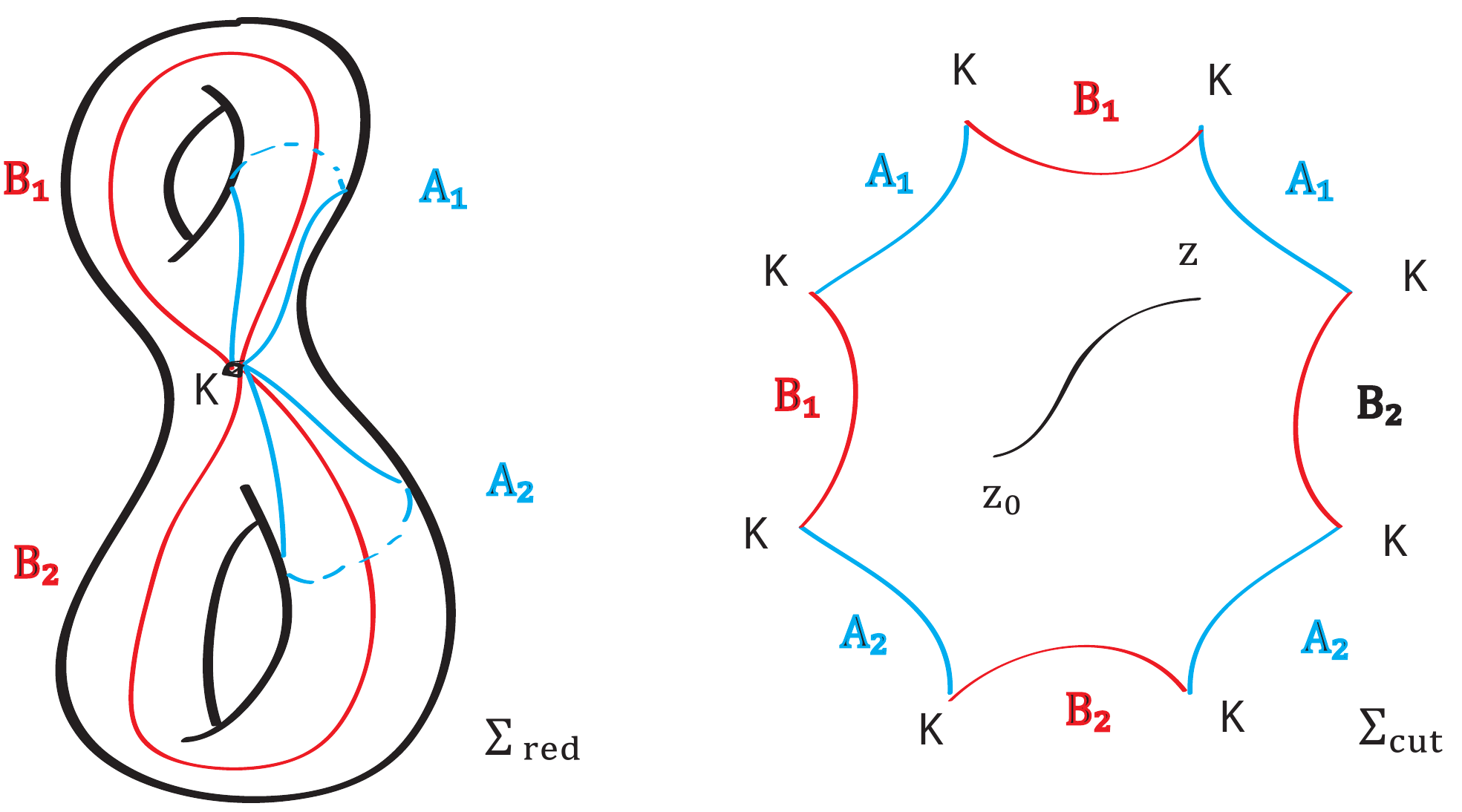}
\end{tabular}
\caption{A genus 2 surface $\Sigma _{\rm red}$ (left) cut into a simply connected domain $\Sigma_{\rm cut}$ in $\CC$ (right).}
\label{fig1}
\end{figure}

Part of the function theory on $\Sigma _{\rm red}$ that we shall need in the presence of R-punctures will be holomorphic and meromorphic sections of the bundles $\cR^{-1}$ and $K \otimes \cR$. 
Generically, on the moduli space for $\Sigma_{\rm red}$ with $2r$ Ramond punctures, we have,
\bea
H^0(\Sigma_{\rm red},{\cal R}^{-1})=r,
\hskip 1in
H^0(\Sigma_{\rm red}, K\otimes {\cal R})=0.
\eea
It is shown in \cite{new.Witten} that these relations hold throughout moduli space when $r\geq g$. To obtain the corresponding holomorphic and meromorphic  sections in terms of $\tet$-functions, it will be necessary to consider the Abel map evaluated on Ramond divisors, such as, for example, 
\bea
\half \int _{\cF_q} ^{\cF_p} \omega _I = \sum _{\eta =1 } ^r \half \int ^{p_\eta } _{q_\eta} \omega _I
\eea
The factors of $\half$ arise in taking the square root of $\cR^{-2}$. Using a description of the surface $\Sigma _{\rm red}$ in terms of the simply connected cut domain $\Sigma_{\rm cut}$, the range of the punctures $p_\eta$ and $q_\eta$ is restricted to $\Sigma _{\rm cut}$. To describe the bundle $\cR^{-1}$ completely and uniquely, including the specification of its generalized spin structure, the Abel map  needs to supplemented by one of the $2^{2g}$ half-integer characteristics, which we shall denote by  $\delta$ below. It is in this sense that $\delta$, for a given choice of the cut domain $\Sigma_{\rm cut}$, labels all the generalized spin structures.

\sm

Explicit expressions for the $r$ holomorphic sections of ${\cal R}^{-1}$ will  be obtained in the remaining parts of this section. Since the case of two R-punctures is illuminating by itself, we shall treat it first, before moving on to the case of general $r \geq 1$.

\subsection{Spinors and propagators for 2 Ramond punctures}

The explicit formulas for the holomorphic and meromorphic sections of $\cR^{-1}$  are most transparent in the simplest case of two R-punctures, denoted $p$ and $q$, with $\cF=p+q$, and we shall discuss this case first. The holomorphic section $h(z)$ for generic moduli  is given by,
\bea
\label{h}
h(z)
=
{\tet[\delta](z-{1\over 2}p-{1\over 2}q)\over\tet[\delta]({1\over 2}p-{1\over 2}q)}
\left ( {E(p,q)\over E(z,p)E(z,q)} \right )^{1\over 2}.
\eea
The expression (\ref{h}) should be understood as giving the section $h(z)$ on $\Sigma_{\rm red}\setminus {\cal F}$. The expression for $h(z)$ in a coordinate chart centered at either one of the punctures $p$ or $q$ is obtained by multiplying (\ref{h}) by the suitable transition function for the bundle ${\cal R}^{-1}$. In particular, in such a chart, $h(z)$ is indeed given by a holomorphic function near $p$ or $q$. 

\sm

The expression (\ref{h}) shows explicitly that $h(z)$ is non-trivial only  for generic moduli, such that $\tet[\delta](z-{1\over 2}p-{1\over 2}q)$ does not vanish identically in $z$, or equivalently, that $\delta - {1\over 2}p-{1\over 2}q$ not be a {\sl special divisor}. It is also instructive to determine how $h(z)$ transforms as $p$ is moved around an arbitrary full period $2\ep$. Because of the presence of $p/2$ in the argument of the $\tet$-function in (\ref{h}), 
the monodromy of the Abel map as $p$ is moved around an arbitrary full period $2 \ep$ 
will produce a shift by an arbitrary  half-period $\ep$ in the argument of the $\tet$-function.
This will  map the holomorphic section $h $ with generalized spin structure $\delta$ to a holomorphic
section $h$ with generalized spin structure $\delta + \ep$ thus producing the transitive 
action on all spin structures, even and odd, as reviewed in \S 4.2.4 of \cite{Witten:2012ga}.

\sm

The other ingredient that we need is a worldsheet fermion propagator, or Szeg\"o kernel,  $S(z,w)$ for spinors on a surface with R-punctures. It is a meromorphic section of ${\cal R}_z^{-1}\otimes{\cal R}_w^{-1}$, and is characterized by the following conditions
\bea
\label{pole}
\pbz S(z,w)=2\pi \delta(z,w),
\eea
together with the anti-symmetry requirement
\bea
\label{antisymmetry}
S(z,w)=-S(w,z).
\eea
Note that the condition (\ref{pole}) alone cannot completely specify $S(z,w)$, because of the existence of the holomorphic spinor $h(z)$. To construct $S(z,w)$, we begin by considering the propagator $S^{pq}(z,w)$, also valued in ${\cal R}_z^{-1}\otimes{\cal R}_w^{-1}$ and defined explicitly by
\bea
\label{1e6}
S^{pq}(z,w)
=
{\tet[\delta](z-w+{1\over 2}p-{1\over 2}q)
\over
\tet [\delta](\half p - \half q ) E(z,w)} \left ( {E(z,p)E(w,q)\over E(w,p)E(z,q)} \right )^{1\over 2}
\eea
The propagator $S^{pq}(z,w)$ satisfies the condition (\ref{pole}), together with
\bea
\label{1e6a}
S^{pq}(p,w)=0,
\hskip 1in
S^{pq}(z,q)=0.
\eea
The propagator $S^{pq}(z,w)$ is not anti-symmetric in $z$ and $w$, and we have instead
\bea
\label{1e7}
S^{pq}  (z,w) + S^{pq} (w,z) = h  (z) h (w)
\eea
As a result, the unique Szeg\"o kernel $S(z,w)$ which is anti-symmetric under the 
interchange of its arguments is defined by
\bea
\label{1e8}
S (z,w) = S ^{pq}  (z,w) - \half   h  (z) h (w).
\eea
Equivalently, this unique Szeg\"o kernel may be obtained by anti-symmetrizing $S^{pq}(z,w)$,
\bea 
\label{1e9}
S(z,w) = \half S^{pq}(z,w) - \half S^{pq}(w,z)
\eea
Its only pole is at $z=w$, with unit residue so that $S(z,w) = 1/(z-w) +\cO(z-w)$.

\subsection{Field theoretic interpretation}

The preceding function theory on $\Sr$ can be expressed in terms of fermion fields $\psi$ and $\bar\psi$ valued in ${\cal R}^{-1}$ and spin fields $\Sigma^{\pm}$. Let the OPE relations of these fields be given by
\bea
\label{1e1}
\bar \psi (z) \, \psi (w) \sim (z-w)^{-1} & \hskip 0.8in & 
\bar \psi (z) \Sigma ^+ (w) \sim (z-w)^{- \half} \, \Sigma ^-(w)
\no \\
&  & \psi (z) \Sigma ^- (w) \sim (z-w)^{- \half} \, \Sigma ^+(w)
\eea
The OPEs $\psi (z) \psi (w)$ and $\bar \psi (z) \bar \psi (w)$ are $\cO(z-w)$ while the 
OPEs $\psi (z) \Sigma ^+ (w)$ and $\bar \psi (z) \Sigma ^- (w)$ are of order $ \cO (z-w)^{\half}$,
neither of which will be needed in the sequel. Then we have
\bea
\label{1e2}
h(z) =
{\<\psi(z)\Sigma^-(p)\Sigma^-(q)\> \over \<\Sigma^+(p)\Sigma^-(q)\>},
\hskip 0.8in 
S^{pq}(z,w)
=
{\<\psi(z)\bar\psi(w)\Sigma^+(p)\Sigma^-(q)\> \over \<\Sigma^+(p)]\Sigma^-(q)\>}.
\eea
The OPE relations guarantee that the relations (\ref{1e6a}) hold automatically. The field theoretic formulation of the anti-symmetric $S$ is obtained by combining (\ref{1e9}) with (\ref{1e2}).

\subsection{The general case of $2r$ Ramond punctures}

We extend  all the preceding concepts to the general case of $2r$ R-punctures. Recall that the $2r$ punctures may be grouped into pairs $(p_\eta, q_\eta)$ with $\eta =1,\cdots, r$. 

\subsubsection{Holomorphic sections}

For generic moduli, the space $H^0(\Sr,{\cal R}^{-1})$ has dimension $r$. An explicit basis for its sections is given by the following formula, valid on $\Sr\setminus \cF$,
\bea
\label{3e}
h^p _\eta  (z) = 
{ \tet [\delta ] (z - p_\eta + \half \cF_p - \half \cF_q) E(p_\eta, q_\eta) ^\half \over 
	\tet [\delta ] (\half \cF_p - \half \cF_q) E(z,p_\eta)^\half E(z,q_\eta)^\half  }
 \prod _{\g \not = \eta} \left ( {E(z,p_\g)  E(p_\eta, q_\g)  \over E(z,q_\g) E(p_\eta, p_\g)} \right )^\half
\eea
Its field theoretic interpretation is
\bea
\label{3f}
 h ^p_\eta (z) = { \< \psi (z) 
\Sigma ^+(p_1) \cdots \Sigma ^+(p_{\eta -1})
\Sigma ^-(p_\eta ) 
\Sigma ^+(p_{\eta+1}) \cdots \Sigma ^+(p_r) 
\Sigma ^-(q_1) \cdots \Sigma ^- (q_r)\> \over 
 \<  \Sigma ^+(p_1) \cdots  \Sigma ^+(p_r)  \Sigma ^-(q_1) \cdots \Sigma ^- (q_r) \>}
\eea
The normalization is arranged so that the monodromy of $h_\eta^p (z)$, 
as the R-punctures $p_\eta ,q_\eta $ are moved around $\Sr$, is just factors of $\pm 1$ or $\pm i$.
The normalizations of the OPE relation $\psi (z) \Sigma ^-(p_\eta ) \sim (z-p_\eta )^{-\half} \Sigma ^+(p_\eta )$ 
and of the behavior of $h^p_\eta  (z)$ in $z$ at the point $p_\eta$ implied by (\ref{3e})
precisely match, and we have $(z-p_{\eta '} )^\half h^p_\eta  (z) \to \delta _{\eta \eta '}$ as $z\to p_{\eta'}$.

\sm

It is natural and  useful to introduce also a conjugate basis 
of holomorphic sections of $\cR^{-1}$  by effecting the interchange of 
$p_\eta \leftrightarrow q_\eta$ for all $\eta$. We shall denote these sections by 
$h ^q _\eta  (z)$. There exists a linear relation between the $h^p_\eta  $ 
and $h ^q _\eta $, by an invertible matrix $\Lambda$,
\bea
\label{hhprime}
h ^p_\eta  (z) = \sum _\zeta \Lambda _{\eta \zeta} \, h _\zeta ^q (z)
\eea
In terms of $\Lambda$, we may now collect all the limits we shall need, 
\bea
\label{hlimq}
\lim _{z \to p_{\eta '} } (z-p_{\eta '} )^\half h^p_\eta  (z) = \delta _{\eta \eta '}
& \hskip 0.7in &
\lim _{z \to q_{\eta '} } (z-q_{\eta '} )^\half  h^p_\eta  (z) = \Lambda _{\eta \eta '}
\no \\
\lim _{z \to q_{\eta '} } (z-q_{\eta '} )^\half h_\eta^q   (z) = \delta _{\eta \eta '}
& &
\lim _{z \to p_{\eta '} } (z-p_{\eta '} )^\half  h_\eta^q   (z) = (\Lambda^{-1})  _{\eta \eta '}
\eea
We shall show below by indirect arguments that the matrix $\Lambda$ satisfies $\Lambda ^t \Lambda = -I$,
so that the change of basis $h^p \leftrightarrow h^q$ is indeed guaranteed to be invertible.

\subsubsection{The Szeg\"o kernel}

Next, we construct Szeg\"o kernels satisfying the same conditions (\ref{pole}) and (\ref{antisymmetry}) as in the simplest case of 2 R-punctures. Again, the condition (\ref{pole}) is unaffected by the addition of an arbitrary linear combination of $h^p_\eta  (z) h^p _\g (w)$  to $S(z,w)$. As in the case of $2$ R-punctures, we begin by considering the following Szeg\"o kernel,
\bea
S^{pq} (z,w) & = & 
{\<\psi(z)\, \overline{\psi(w)} \, \Sigma^+(p_1) \cdots \Sigma ^+ (p_r)\, 
\Sigma ^- (q_1 ) \cdots \Sigma^-(q_r)\>
\over \< \Sigma^+(p_1) \cdots \Sigma ^+ (p_r) \Sigma ^- (q_1 ) \cdots \Sigma^-(q_r) \>}
\eea
By construction, we have,
\bea
\label{3S0}
S^{pq}  (p_\eta ,w)= S^{pq}  (z,q_\eta)=0 \hskip 1in \eta =1,\cdots, r
\eea
On $\Sr \setminus \cF$, the Szeg\"o kernel is given by the following explicit formula,
\bea
\label{3b}
S^{pq} (z,w)
=
{\tet[\delta](z-w+ \half \cF_p - \half \cF_q)
\over
\tet [\delta](\half \cF_p - \half \cF_q ) \, E(z,w)} 
\prod _\g \left ( {E(z,p_\g )E(w,q_\g )\over E(w,p_\g )E(z,q_\g )} \right )^{1\over 2}
\eea
Since the Szeg\"o kernel $S^{pq} (z,w)$ is not anti-symmetric in $z,w$, we consider its symmetric part $S^{pq}(z,w)+S^{pq}(w,z)$. This sum is a holomorphic section of $\cR^{-1}_z \otimes \cR^{-1}_w$ and may be written in two equivalent ways, either in the basis generated by $h^p_\eta $, or in the basis generated by $h^q _\eta $, 
\bea
\label{3S1}
S^{pq}  (z,w) + S^{pq} (w,z) =
\sum _{\eta , \eta' } H^p _{\eta \eta'} h^p_\eta   (z) h^p_{\eta'}  (w)
=
\sum _{\eta , \eta' } H^q _{\eta \eta'} h_\eta ^q  (z) h_{\eta'} ^q (w)
\eea
By construction, the matrices $H^p _{\eta \eta'}$ and $H _{\eta \eta'} ^q $ are symmetric in $\eta, \eta'$. They may be determined using  the relations of (\ref{3S0}), namely $S^{pq}  (p_\eta,w)=0$ on the first line, and $ S^{pq} (z,q_\eta)=0$ on the second line, along with the following limits
\bea
\label{Slim}
\lim _{w \to p_\eta } (w-p_\eta )^\half S^{pq}   (z,w) & = & + h^p_\eta   (z)
\no \\
\lim _{z \to q_\eta} (z-q_\eta)^\half S^{pq} (z,w) & = & -  h^q_\eta  (w)
\eea
which may be read off from the  explicit expressions given for $S^{pq} $, $h^p _\eta $, and $h^q _\eta $,
Applying these limits to both sides of (\ref{3S1}) and using (\ref{Slim}), we obtain,
\bea
\label{HH}
H^p_{ \eta \eta '} = + \delta _{\eta \eta'} \hskip 1in H^q_{\eta \eta'} = - \delta _{\eta \eta'}
\eea
a result which is in agreement with the case  $r=1$ computed previously. Using the second equality in (\ref{3S1}), along with the equations of (\ref{HH}) and the linear relation between $h^p _\eta $ and $ h _\eta ^q $ in (\ref{hhprime}), we deduce the previously announced result,  
\bea
\label{lamsq}
\Lambda ^t \Lambda = - I
\eea 

\sm

Returning to the construction of an anti-symmetric Szeg\"o kernel, we introduce the following family $S^M$ of candidate Szeg\"o kernels, indexed by an $r\times r$ matrix $M$ of parameters,
\bea
\label{3S2}
S^M  (z,w) = 
S ^{pq} (z,w) + \sum _{\eta \eta'} \half \Big ( M_{\eta \eta'} - \delta_{\eta \eta'} \Big )
h^p_\eta   (z) h^p_{\eta'}  (w).
\eea
We have the following reflection symmetry relation, 
\bea
S^M (w,z) = - S^{( -M^t ) } (z,w)
\eea
where $M^t$ is the transpose of $M$. A Szeg\"o kernel  satisfying (\ref{pole}) and (\ref{antisymmetry}) can then be obtained by setting $M$ to be any anti-symmetric matrix. Up to this point, the anti-symmetric matrix $M$ is arbitrary. But we shall see later that a unique choice is dictated by the requirement that the ensuing super period matrix be symmetric in the ${\bf Z}_2$-graded sense.

\newpage

\section{Superholomorphic Forms and their Periods}
\setcounter{equation}{0}
\label{sec3}

The key ingredients in the construction of the period matrix for an ordinary Riemann surface were the holomorphic 1-forms $\om_I$ which are holomorphic sections of the canonical bundle $K$. Analogously, in the supergravity formulation, the key ingredient for the super period matrix will be the superholomorphic 1/2 forms \cite{D'Hoker:1988ta,D'Hoker:1989ai}.  A formal approach to the super period matrix based more on algebraic geometry may be found in \cite{RSV}. A useful review of the subject may be found in  \cite{Witten:2012ga}. In this section, we shall give a description and explicit construction of these forms for super Riemann surfaces of general genus $g$ and with an arbitrary number $2r$ of R-punctures. We shall then use those forms to derive even and odd periods.

\subsection{Differential equations for superholomorphic 1/2 forms}

We shall be interested in superholomorphic forms $\hat \omega$ of $U(1)$ weight $1/2$ and their periods for a given super Riemann surface defined in the supergravity formulation by the equivalence class $(\gg,\chi)$. The condition for superholomorphicity reads $\cD_- \hat \om=0$, where $\cD_-$ is the analog of the Cauchy-Riemann operator for the super Riemann surface $\Sigma$ (see for example \cite{D'Hoker:1988ta}). 

\sm

We introduce a system of local complex coordinates $(z, \bar z)$ associated with the complex structure given by the metric $\gg$, and an odd coordinate $\theta$ which should be viewed as a section of $\cR$. The forms $\hat \om$ may be decomposed $\hat\omega (z,\theta ) =\hat\omega_+(z)  + \theta\,\hat\omega_z (z) $ in components $\hat \om _+$ and $\hat \om _z$ which are sections of $\cR^{-1}$ and $K$ respectively. The condition for superholomorphicity $\cD_- \hat \om=0$ is then given by the following component equations,\footnote{The effects of auxiliary fields, which are present in the superfields $\hat \om$ and the supergravity superfields associated with $(\gg,\chi)$, may be consistently ignored in the process, as was shown in detail in \cite{D'Hoker:1989ai}.}  
\bea
\label{superholomorphic}
&&
\p_{\bar z} \hat \omega_+ +{1\over 2}\chi \, \hat \omega_z=0
\nonumber\\
&&
\p_{\bar z}\hat\omega_z +{1\over 2}\p_z(\chi \, \hat \omega_+)=0.
\eea
As a result, the 1-form $dz\,\hat\omega_z -{1\over 2}d\bar z \,\chi \, \hat\omega_+ $ is closed, and we can define the period of $\hat\omega$ around a closed cycle $C$ by 
\bea
\label{period}
\oint_C\hat\omega=
\oint_C \left ( dz\,\hat\omega_z -{1\over 2}d\bar z\,\chi \, \hat\omega_+ \right ).
\eea
We shall see that in the presence of R-punctures, superholomorphic forms of weight 1/2 exist for both even or odd grading. The components $\hat\omega_z$ in the odd forms are even, and their periods along closed contours $C$ are even moduli. The components $\hat \omega_z$ in the even forms are odd, and their periods along closed contours $C$ are odd moduli.

\sm

Considering now the generalized situation in which we view a super Riemann surface as a deformation,
parametrized by $(\chi, \mu)$ of a reduced Riemann surface $\Sr$ with metric $\mg$, we may write down also the equations for superholomorphicity $\cD_- \hat \om=0$ of a form $\hat\omega$ in this context. 
We now introduce a system of local complex coordinates $(z,\bar z)$ associated with the metric $\mg$ of $\Sr$, and carry out the deformation of complex structures by $\mu$ on equations (\ref{superholomorphic}) by replacing $\p_{\bar z} \to \p_{\bar z} + \mu \p_z$ on scalars, with analogous modifications for spinors and tensors.  The $\cD_- \hat \om=0$ equations then read as follows,
\bea
\label{1c2}
\left ( \pbz + \mu \, \pz + \half (\pz \mu) \right ) \hat \om _+ + \half \chi \, \hat \om _z & = & 0
\no \\
\pbz \hat \om_z + \p_z \left ( \half \chi \, \hat \om_+ + \mu \, \hat \om _z \right ) & = & 0
\eea 
and the periods of a superholomorphic form $\hat\omega$ receive a $\mu$-dependent correction, 
\bea
\label{period1}
\oint_C\hat\omega
=
\oint_C  \left ( dz\, \hat\omega_z-d\bar z\, \mu \, \hat\omega_z - {1\over 2}d \bar z \, \chi \, \omega_+  \right ) 
\eea
The equations of (\ref{1c2}) are invariant under conformal transformations with
$\hat \om_+, \hat \om_z, \chi$, and $\mu$ transforming respectively with weights
$(1/2, 0), (1,0), (-1/2,1)$, and $ (-1,1)$. They are also invariant under local supersymmetry
transformations with parameter $\xi $, provided $\chi$ and $\mu$ are transformed as in (\ref{susy}) and the  components of $\hat \om$ transform as follows,
\bea
\label{susy1}
\delta _\xi \hat \om _+ = \xi \, \hat \om_z 
 \hskip 1in  
\delta _\xi \hat \om _z = \p_z \left ( \xi \, \hat \om _+ \right )
\eea

\subsection{A basis of superholomorphic forms of U(1) weight $1/2$}

Using the function theory on $\Sr$ described in \S 2, we can easily write down a basis $(\hat\omega_I^M|\hat\omega_\eta^M)$ of superholomorphic forms of U(1) weight $1/2$. The upper index $M$ refers to the choice of propagators $S^M(z,w)$ in the expressions below.
First, since $\chi$ and $\mu$ are nilpotent, it is advantageous to rewrite the equation (\ref{1c2}) for superholomorphicity as an integral equation which can be solved after a finite number of iterations,
\bea
\label{1d1}
\hat \om _+ ^M(z) & = & \varpi _+ (z) 
- { 1 \over 2 \pi} \int_\Sr \!\!\! d^2w \,  S^M(z,w) 
\left ( \half  \chi  \, \hat \om _w^M + \mu \, \p_w \hat \om _+^M + \half (\p_w \mu) \, \hat \om_+^M \right )(w)
\no \\
\hat \om _z ^M(z) & = & \varpi _z (z) 
+ {1 \over 2 \pi} \int _\Sr \!\!\! d^2w \, \om(z,w) 
\left (  \mu \, \hat \om _w^M + \half \chi \, \hat \om_+^M \right ) (w)
\eea
The contributions $\varpi _+ (z) $ and $\varpi _z (z) $ are respectively any holomorphic half-form and any holomorphic one-form with respect to the complex structure for $\Sr$.  In the above integral equations, we have further used the inverse of $\pbz$ on one-forms and half-forms.  The inverse of the operator $\pbz$ on 1-forms is proportional to $\om (z,w) = \pz \pw \ln E(z,w)$. The inverse of the operator $\pbz $ on sections of $\cR^{-1}$ is given by the Szeg\"o kernel $S^M(z,w)$ in the presence of R-punctures described in \S 2. Different choices of matrices $M$ in the Szeg\"o kernel $S^M(z,w)$ will result in different choices of basis vectors for the same space of superholomorphic forms.

\sm

For the odd superholomorphic 1/2 forms, we choose $\varpi _z  =\omega_I$ and $\varpi _+ =0$, where $\omega_I$ is the basis of holomorphic forms for $\Sr$, with the normalization (\ref{Anorm}), and we obtain the following basis of $g$ odd superholomorphic 1/2 forms, 
\bea
\label{1h5}
\hat \om _{I+}^M (z) & = &
- { 1 \over 2 \pi} \int_\Sr \!\!\! d^2w \, S^M(z,w) 
\left ( \half  \chi  \, \hat \om _{Iw}^M + \mu \, \p_w \hat \om _{I+} ^M+ \half (\p_w \mu) \, \hat \om_{I+}^M \right ) (w)
\no \\
\hat \om _{Iz}^M (z) & = & \omega _I(z) +
 {1 \over 2 \pi} \int _\Sr \!\!\! d^2w \, \om (z,w) \left (  \mu \, \hat \om _{Iw} ^M+ \half \chi \, \hat \om_{I+} ^M\right ) (w)
\eea
The superscript $M$ shows the dependence of the forms on the as yet unspecified matrix $M$.

\sm

For the even superholomorphic 1/2 forms, we choose $\varpi_z =0$ and $\varpi _+ =h_\eta^N$, where $h^N_\eta$ is an as yet unspecified linear combination of the holomorphic section $h^p_\eta$ of $\cR^{-1}$ given in \S 2. This linear combination may be expressed in terms of an as yet unspecified  $r \times r$ matrix $N$,
\bea
\label{1h3}
h^N _\eta (z) = \sum _\a N_{\eta \a} h^p _\eta (z)
\eea
The resulting even superholomorphic 1/2 forms will be denoted $\rh_\eta^{MN}$ thereby exhibiting their dependence on the matrices $M,N$. They are given by
\bea
\label{1h4}
\rh _{\eta+}^{MN} (z) & = &
h^N_\eta (z) 
- { 1 \over 2 \pi} \int_\Sr \!\!\! d^2 w \, S^M(z,w) 
\left ( \half  \chi  \, \rh _{\eta w}^{MN} + \mu \, \p_w \rh _{\eta +} ^{MN}
+ \half (\p_w \mu) \, \rh_{\eta +} ^{MN}\right ) (w)
\no \\
\rh _{\eta z}^{MN}(z) & = & 
 {1 \over 2 \pi} \int _\Sr \!\!\! d^2w \,  \om (z,w) \left (  \mu \, \rh _{\eta w}^{MN} + \half \chi \, \rh_{\eta +} ^{MN}\right ) (w)
\eea
We have obtained in this way a $g|r$-dimensional basis of superholomorphic forms $(\hat\omega_I^M|\rh_\eta^{MN})$, with $I=1,\cdots, g$ and $\eta = 1, \cdots,  r$, generated from their values at $\chi=0$, $\mu=0$ by a finite number of iterations when $\chi$ and $\mu$ are generated by a finite number of odd moduli. 

\sm

The evaluation of the periods of the basis $(\hat\omega_I^M |\rh_\eta^{MN} )$ is a bit subtle, because the expressions in (\ref{1h5}) involve  integrals that are only conditionally convergent. We shall discuss their evaluation in detail below in \S 3.7, but for the moment, we shall quote the result that the basis does have the expected $A$-periods,
\bea
\oint_{A_J}\hat\omega_I^M=\delta_{IJ},
\hskip 1in
\oint_{A_J}\rh_\eta^{MN}=0.
\eea
There is still some arbitrariness in this basis. The forms $\rh_\eta^{MN}$ have vanishing $A$-periods, so that the normalization of $A$-periods for $\hat\omega_I^M$ can only determine these forms up to linear combinations with odd coefficients of the forms $\rh_\eta^{MN}$. This is related to the arbitrariness in the choice of anti-symmetric matrix $M$ in the Szeg\"o kernel $S^M(z,w)$. Also, the normalization of the forms $\rh^{MN}_\eta$ has not yet been fixed, but will determine the matrix $N$ uniquely.

\subsection{Witten's basis and Super Period Matrix}

A key insight of \cite{new.Witten} is that the residual arbitrariness in the basis of superholomorphic forms requires a proper normalization, defined by introducing so-called ``fermionic periods". Once this proper normalization is satisfied, the resulting matrix of periods will have all the desirable properties, including symmetry with respect to the natural ${\bf Z}_2$ grading.

\sm

The fermionic periods of a superholomorphic form can be constructed as follows. Let $\hat\omega(z,\theta)=\hat\omega_+(z)+\theta\hat\omega_z(z)$ denote any even or odd superholomorphic form of $U(1)$ weight $1/2$. The definite integral of $\hat\omega$ between two points $(z,\theta_z)$ and $(w,\theta_w)$ in $\Sigma$ is given by
\bea
\label{integ}
\int_{w,\theta_w}^{z,\theta_z}\hat
\omega
=
\int_w^z \left ( dz\, \hat\omega_z-d\bar z\, \mu \, \hat\omega_z -{1\over 2}d \bar z \, \chi \, \omega_+ \right )
+
\theta_z\hat\omega_+(z)-\theta_w\hat\omega_+(w)
\eea
(see e.g. eq. (5.13) of \cite{D'Hoker:1989ai}). It is natural to introduce an associated 1-form following \cite{Witten:2012ga}, 
\bea
\kappa (z, \theta) = dz\, \hat \omega_z(z) -  d\bar z \, \mu (z) \hat\omega_z (z) 
-{1\over 2} d \bar z \, \chi (z) \omega_+  (z) +d(\theta\hat\omega_+(z)).
\eea
The form $\kappa$ is closed since its components satisfy (\ref{1c2}).
Next, use local complex  coordinates $(z|\theta)$ near the R-puncture $p_\eta$ for which $z=0$ at $p_\eta$,
and in which  the superconformal structure is defined by the distribution generated by 
\bea
\label{dist}
D_\theta ^* = { \p \over \p \theta } + \theta z { \p \over \p z}
\eea
The odd period $w^p  _\eta (\hat \om)$ of the closed form $\kappa$ which is related to the superholomorphic
1/2 form $\hat \om$ at the puncture $p_\eta$ is defined in \cite{new.Witten} by
\bea
\label{mN}
\kappa = {1 \over \mN} \, w^p _\eta(\hat \om) \, d \theta  \quad {\rm mod} ~ z
\hskip 1in 
\mN = \sqrt{2 \pi \sqrt{-1}}
\eea
The fermionic period $w^p  _\eta(\hat \om)$ must be independent of $\theta$ since $d \kappa =0$. The purpose of the  normalization factor $\mN$  is to assure symmetry of the super period matrix, in the $\ZZ_2$ graded sense. Similarly, one defines the fermionic period $w^q  _\eta$ at the puncture $q_\eta$. The fermionic periods $w_\eta^A(\hat\omega)$ and $w_\eta^B(\hat\omega)$ (denoted $w_\eta (\kappa)$ and $\tilde w ^\eta(\kappa)$ in \cite{new.Witten}) can now be defined by
\bea
\label{oddper}
w_\eta^A (\hat\omega) & = &  
{1 \over \sqrt{2} } \left ( w^p  _\eta (\hat\omega) + \sqrt{-1} \, w^q _\eta (\hat\omega) \right )
\no \\
w_\eta^B (\hat\omega) & = &  
{1 \over \sqrt{2} } \left ( w^p _\eta (\hat\omega) - \sqrt{-1} \, w^q _\eta (\hat\omega) \right ).
\eea
Witten's normalization prescription for a basis of closed holomorphic 1-forms $(\sigma _I | \nu_\eta)$ with $I=1,\cdots, g$ and $\eta=1,\cdots, r$ for a super Riemann surface with genus $g$ and $2r$ punctures is then
given as follows,
\bea
\label{normper}
\oint _{A_J} \sigma _I = \delta _{IJ} 
& \hskip 1in &
w_\eta^A(\sigma _I) =0 
\no \\
\oint _{A_I} \nu_\eta  =0 \hskip 0.15in 
&&
w_\eta^A (\nu_\zeta)=\delta_{\zeta\eta}
\eea
Witten's $ g|r \, \times \, g|r$ super period matrix can now be defined by\footnote{Possible confusion on the values taken by the indices $I$ and $\eta$ may be avoided by denoting these values by distinct symbols, $I=1,2, \cdots, g$ and $\eta = \tilde 1, \tilde 2, \cdots, \tilde r$, and interpreting the numbers with tildes as ``fermionic".}
\bea
\hat \Omega = \pmatrix{\hat\Omega_{IJ} & \hat\Omega_{I\zeta}\cr
\hat\Omega_{\eta J}&\hat\Omega_{\eta\zeta}\cr}
\eea
with the entries given by
\bea
\oint _{B_J} \sigma _I = \hat \Omega _{IJ} 
& \hskip 1in &
w_\eta^B(\sigma _I) =\hat \Omega _{I \eta}  
\no \\
\oint _{B_I} \nu _\eta = \hat \Omega _{\eta I}
&&
w_\eta^B (\nu_\zeta)= \hat \Omega_{\zeta \eta}.
\eea
It was shown in \cite{new.Witten} that the normalization conditions (\ref{normper}) can be implemented in the complement of a divisor in the reduced space of $\mM_{g,0,2r}$ and that, using the Riemann relations, they imply symmetry in the ${\bf Z}_2$ graded sense of the matrix $\hat \Omega$ so that its components satisfy 
$\hat \Omega_{IJ}=\hat\Omega_{JI}$, $\hat \Omega _{I\eta} = \hat \Omega _{\eta I}$, and $\hat \Omega _{\eta \zeta}=- \hat\Omega_{\zeta\eta}$.

\subsection{Witten's basis in the supergravity formalism}

We shall now show how to implement Witten's normalization explicitly in terms of the superholomorphic  1/2 forms constructed in \S 3.2. The issue is to understand and evaluate the change of bases,
\bea
(\hat\omega_I^M|\rh_\eta^{MN})\ \leftrightarrow\ (\sigma_I|\nu_\eta).
\eea
Recall that the basis $(\hat\omega_I^M|\rh_\eta^{MN})$ depends on the choice of Szeg\"o kernel $S^M(z,w)$ for $r>1$, a choice which is parametrized by the matrix $M$, as well as on the choice of the basis $h_\eta^N$ as a function of $h^p_\eta$ of (\ref{1h3}), a choice which is parametrized by the matrix $N$. It will turn out that the matching of the basis $(\hat\omega_I^M)\leftrightarrow (\sigma_I)$ of odd forms corresponds to a unique choice of the anti-symmetric matrix $M$ in $S^M(z,w)$. Once $M$ has been determined, the matching of the basis $(\rh_\eta^{MN})\leftrightarrow (\nu_\eta)$ of even forms is obtained by enforcing the fermionic $A$-periods of $\rh_\eta^{MN}$ and will uniquely determine the matrix $N$.

\subsubsection{The case of $2$ Ramond punctures}

In this case, there is a unique Szeg\"o kernel $S(z,w)$ which is anti-symmetric, so the basis $(\hat\omega_I|\rh_0)$ obtained in \S 3.2 is canonical, up to an overall normalization of $\rh _0$.  We shall indeed show that it essentially coincides with Witten's basis $(\sigma _I |\nu_0)$
\bea
(\hat\omega_I|\rh_0) \leftrightarrow
(\sigma_I| \sqrt{2} \, \mN \, \nu_0)
\eea
where $\mN$ is the normalization given in (\ref{mN}). Here and below when discussing the $r=1$ case,  we shall use the notation 0 for the index $\eta$ in order to clearly distinguish it from the index $I$ which takes values $I=1,\cdots, g$.

\sm

The 1-form $\nu_0$ corresponds to a superholomorphic 1/2 form proportional to $\rh_0$, 
since both are required to have vanishing $A_I$-periods. The constant of proportionality
may be computed from the explicit expression for the components of $\rh_0$ in (\ref{1h4}),
the normalization of $h$ in (\ref{h}), and the normalization $w^A(\nu_0)=1$.

\sm

To evaluate $w^A(\rh_0)$, we must evaluate $w_p(\rh_0)$ and $w_q(\rh_0)$.
Near $p$ we use coordinates $(z|\theta)$ in which $D^*_\theta$ takes the canonical form (\ref{dist}).
The formulas for $h(z)$ and $S ^{pq}(z,w)$ derived in (\ref{h}) and (\ref{1e6}) respectively
are expressed instead in terms of coordinates $(z| \hat \theta)$ in which $z-p=0$ at $p$, and in 
which the distribution takes the form $D^* _\theta = \p_{\hat \theta} + \hat \theta \p_z$ away from $\cF$. The map between the coordinate systems is provided by $\hat \theta_p = (z-p)^\half \theta_p$. Similarly near $q$,
the coordinates map via $\hat \theta_q = (z-q)^\half \theta_q$. This map allows us to work out the 
odd periods $w_p (\rh_0)$, since near $p$ the corresponding 1-form is  given by
\bea
\kappa= \rh_{0+}(z) d \hat \theta_p = (z-p)^\half \rh_{0+}(z) d \theta_p \quad {\rm mod} ~ (z-p)
\eea
We then have
\bea
w^p (\rh_0) & = & \mN  \, \lim _{z \to p} \left ( (z-p)^\half \rh_{0+} (z) \right )
\no \\
w^q (\rh_0) & = & \mN \, \lim _{z \to q} \left ( (z-q)^\half \rh_{0+} (z) \right )
\eea
To compute these limits, we evaluate the contribution from $h (z)$ as well as the contribution from the Szeg\"o kernel $S (z,w)$ of (\ref{1e8}). The first, near $p$ and near $q$, is given by,
\bea
\label{hlim}
\lim _{z \to p}  (z-p)^\half h(z)   & = & 1
\no \\
\lim _{z \to q}  (z-q)^\half h(z) & = & - \sqrt{-1}
\eea
The factor $\sqrt{-1}$ arises from the square root involving the prime forms as $z \to q$ in (\ref{h}). The signs have been chosen for consistency with the sign conventions of (\ref{normper}). 

\sm

It remains to evaluate the limits for the Szeg\"o kernel $S (z,w)$. This is straightforward with the help of the vanishing conditions (\ref{1e6a}), namely $S^{pq}(p,w) = S^{pq}(z,q)=0$, and the relation (\ref{1e8}).  We find, 
\bea
\label{3d3}
\lim _{z \to p}  (z-p)^\half S(z,w) & = & - \half h (w)
\no \\
\lim _{z \to q}  (z-q)^\half S(z,w) & = & - \half \sqrt{-1} \,  h (w)
\eea
Substituting these results into (\ref{1h4}) for $\eta=0$, we find, 
\bea
\label{wpq0}
w^p (\rh_0) =\mN \hskip 0.5in & \hskip 1in  & w^A (\rh_0) =\sqrt{2} \, \mN
\no \\
w^q (\rh_0) = - \sqrt{-1} \, \mN & \hskip 1in  &  w^B (\rh_0) =0
\eea
As a result, we may identify the one form $ \sqrt{2} \, \mN \,\nu_0$ with the superholomorphic 1/2 form $  \rh_0$.

\sm

Next, we compare $\sigma _I$ and $\hat \omega _I$. We note that their bosonic periods on $A_I$-cycles coincide. Hence, the normalizations of $\sigma _I$ and $\hat \omega _I$ differ at most by the addition a 
multiple of $\rh_0$ to $\hat \om_I$. Since all $\hat \om _I $ with these normalizations differ 
by multiples of $\rh_0$, we shall evaluate the fermionic periods of $\hat \om _I $.
To do so, we now need the limits, 
\bea
\label{wpq}
w^p (\hat \om _I) & = & \mN \, \lim _{z \to p} \left ( (z-p)^\half \hat \om _{I+} (z) \right )
\no \\
w^q (\hat \om _I ) & = & \mN \, \lim _{z \to q} \left ( (z-q)^\half \hat \om _{I+} (z) \right )
\eea
It will be convenient to relate these expressions to the bosonic $B$-periods of $\nu_0$, namely,
\bea
\hat \Omega _{0I} = \oint _{B_I} \nu_0 = {1 \over \sqrt{2} \, \mN} \oint _{B_I} \rh_0
\eea
which may be evaluated with the help of the formula (\ref{B10}) for the $B_I$-period of the Abelian differential of the second kind $\om(z,w)$.
By substituting the limits for the Szeg\"o kernel of (\ref{3d3}) into  (\ref{1h5}). we find, 
\bea
w^p (\hat \om _I) = {  \mN^2 \over 2 \sqrt{2}  \pi i} \, \hat \Omega _{0I},
\hskip 1in 
w^q (\hat \om _I ) =   {  \mN^2 \over 2 \sqrt{2}  \pi i} \sqrt{-1} \, \hat \Omega _{0I}
\eea
As a result, the fermionic periods are given as follows, 
\bea
w^A (\hat \om _I) = 0,
\hskip 1in
w^B (\hat \om _I) = \hat \Omega _{I0}=  { \mN^2  \over 2 \pi  i} \, \hat \Omega _{0I}.
\eea
Proper $\ZZ_2$ graded symmetry of the super period, which includes the relation $\hat \Omega _{I0}=\hat \Omega_{0I}$  requires us to identify the $\sqrt{-1}$ in the definition of $\mN$ with the factor of $i$ in the first line integral of (\ref{B10}). The basis of superholomorphic forms $\hat\omega_I$ obeys Witten's normalization conditions, and therefore must agree with the basis of one forms $\sigma_I$.

\subsubsection{The general case of $2r$ Ramond punctures}

For general numbers $2r$ of R-punctures, the odd and even superholomorphic 1/2 forms are given respectively in (\ref{1h5}) and (\ref{1h4}). They have been normalized only partially, and still involve a dependence on the matrices $M$ and $N$, whose values have not yet been prescribed. A change in the assignment of $M$ will induce a change of basis on $\rh_\eta^{MN}$ without admixture of the forms $\hat \om_I$, while on $\hat \om_I$ the change will act by adding a linear combination of $\rh_\eta^{MN}$. A change in $N$ will only affect $\rh^{MN}_\eta$ by a change of basis. We will show below that, generically, there exists a unique preferred choice of $M$ and $N$, such that 
\bea
\label{generalr}
(\hat\omega_I^M | \rh _\eta^{MN})
\leftrightarrow
(\sigma_I |  \nu_\eta)
\eea
The change of basis is invertible if and only if the matrix $(I + \sqrt{-1} \, \Lambda)$ is invertible, where the matrix $\Lambda$ was defined in (\ref{hhprime}). Thus, in the supergravity formalism, the divisor in the moduli space $\mM_{g,0,2r}$ outside of which Witten's normalization can be implemented is described by the equation
\bea
{\rm det}\,(I+\sqrt{-1}\Lambda)=0
\eea
To prove the above statements, we consider first the normalization of the forms  $\hat \om _I^M$. These forms have already been normalized on bosonic $A$-periods. The fermionic periods of $\hat \om _I ^M$ involve the Szeg\"o kernel $S^M$ used to construct these forms in (\ref{1h5}). We shall now show that, generically, there exists a unique choice of anti-symmetric matrix $M$ which guarantees the vanishing of the fermionic $A$-periods of $\hat \om ^M_I$. 

\sm

To this end, we compute the limits of the Szeg\"o kernel for arbitrary $M$, using the definition (\ref{3S2})
and the relations (\ref{3S0}), and we find, 
\bea
\lim _{z \to p_\eta} (z-p_\eta)^\half S^M (z,w) & = &   
\sum _\b \half \Big ( M_{\eta \b} - \delta _{\eta \b} \Big ) h^p _\b (w)
\no \\
\lim _{z \to q_\eta} (z-q_\eta)^\half S^M (z,w) & = & 
\sum _{\a, \b} \half \Lambda _{\a \eta} \Big ( M_{\a \b} + \delta _{\a \b} \Big )  h^p _\b (w)
\eea
We now form the combinations of these limits that enter into the fermionic periods $w^A_\eta$ and $w^B_\eta$, with respective signs $+\sqrt{-1}$ and $-\sqrt{-1}$ in the formula below, 
\bea
\label{Snorm}
&&
\lim _{z \to p_\eta} (z-p_\eta)^\half S^M (z,w) \pm \sqrt{-1} 
\lim _{z \to q_\eta} (z-q_\eta)^\half  S^M (z,w)
\no \\ && \hskip 1in 
= \sum _\b \half \Big ( M - I \pm \sqrt{-1} \, \Lambda ^t (M+I) \Big ) _{\eta \b} h^p _\b (w)
\eea
For the $w^A_\eta$-period the sign is $+\sqrt{-1}$ and for generic values of $\Lambda$, 
namely again as before when the matrix $(I + \sqrt{-1} \Lambda)$ is invertible, the vanishing
of the right side determines $M$ uniquely to be $M=M_0$, with
\bea
\label{3d9}
M_0= - \Big ( I - \sqrt{-1} \, \Lambda \Big ) \Big ( I + \sqrt{-1} \, \Lambda  \Big )^{-1}
\eea
While not manifest from the expression (\ref{3d9}), the matrix $M_0$ is seen to be anti-symmetric with the help of the relation $\Lambda ^t \Lambda =- I$. In summary, we have $w^A(\hat \om _I ^{M_0})=0$, so that the identification $\sigma _I \leftrightarrow \hat \om_I ^{M_0}$ is complete.

\sm

It remains to normalize the fermionic $A$-periods of $\rh^{M_0 N}_\zeta$. To this end, we again use (\ref{Snorm}) for the $w^A_\eta$ period with the sign $+\sqrt{-1}$, and we find
\bea
w_\eta ^A (\rh^{M_0 N}_\zeta) = \mN \Big ( (I + \sqrt{-1} \, \Lambda) N \Big )  _{\zeta \eta}
\eea
Generically, Witten's normalization of (\ref{normper}) may be achieved by setting $N=N_0$ where 
\bea
N_0 = \mN^{-1} ( I + \sqrt{-1} \, \Lambda )^{-1}
\eea
In summary, we have $w^A_\eta (\rh ^{M_0N_0} _\zeta)= \delta _{\eta \zeta}$, so that the identification $\nu_\eta \leftrightarrow \rh_\eta^{M_0 N_0}$ is complete.

\sm

Henceforth, we shall make the choices $M=M_0$ and $N=N_0$ and drop the corresponding superscripts on the differentials and the superholomorphic 1/2 forms, and simply set
\bea
\label{3d11}
h_\eta(z) \equiv h^{N_0}_\eta (z) \hskip 0.25in
& \hskip 1in &
\hat \om _I (z,\theta) \equiv \hat \om _I^{M_0}(z,\theta)
\no \\
S(z,w) \equiv S^{M_0} (z,w)
&&
\rh_\eta (z,\theta) \equiv \rh ^{M_0 N_0} _\eta (z,\theta)
\eea
The choices for both $M_0$ and $N_0$  are regular provided the matrix $(I +\sqrt{-1} \, \Lambda)$ is invertible
everywhere in the reduced moduli space with R-punctures.

\subsection{Calculation of the bosonic periods of $(\hat\omega_I|\rh_\eta)$}

Our goal is to derive explicit formulas for the periods $\hat\Omega_{IJ}$, $\hat\Omega_{I\eta}$, $\hat\Omega_{\eta I}$, and $\hat\Omega_{\eta\zeta}$ of Witten's basis $(\sigma_I|\nu_\eta)$. Since we have now established the identification $(\hat\omega_I|\rh_\eta) \leftrightarrow (\sigma_I|\nu_\eta)$, it will suffice to determine the periods of $(\hat\omega_I|\rh_\eta)$.

\sm

Recall that the forms $(\hat\omega_I| \rh_\eta)$ were given explicitly by the expressions (\ref{1h5}) and (\ref{1h4}), but these expressions involve conditionally convergent integrals, and the evaluation of their periods has to be carried out with some care. The following formula for the interchange of conditionally convergent integrals was established in \cite{D'Hoker:1989ai}, eq. (5.21). Let $\psi = d\bar z \, \psi _{\bar z}$ be an arbitrary  form of type $(0,1)$, then we have,
\bea
\int_u^v dz
\left (  {1\over 2\pi} \int _\Sr \!\!\! d^2w \, \omega(z,w)\, \psi_{\bar w} \right )
=
\int_u^v d\bar z\, \psi_{\bar z}
+
{1\over 2\pi}
\int _\Sr \!\!\! d^2w \,  \left ( \int_u^v dz\,\omega(z,w)\, \psi_{\bar w} \right ).
\eea
For the integrals of the components of the form $\hat \om _I $ of interest here, the first term on the above right hand side compensates precisely the second term on the right hand side of the expression (\ref{period1}) defining periods of superholomorphic forms. Using this result, we find that the integrals of the forms $\hat \om _I$ along any closed cycle $C$ is given by,
\bea
\oint_C \hat\omega_I
=
\oint_C\omega_I
- { 1 \over 2 \pi} \int  _\Sr \!\!\! d^2w \, \oint_C dz\,\omega(z,w)\, 
\left ( \mu \, \hat \omega_{Iw}+{1\over 2}\chi \,  \hat\omega_{I+} \right ).
\eea
The three integrals we shall need here may be evaluated with the help of (\ref{B10}),
\bea
\label{3e4}
\oint _{A_I} dz \int _\Sr \!\!\! d^2u \,  \om (z,u) \, \psi_{\bar u} - 2 \pi \oint _{A_I} d\bar z \, \psi _{\bar z} & = & 0
\no \\
\oint _{B_I} dz \int _\Sr \!\!\! d^2u \,  \om (z,u) \, \psi_{\bar u} - 2 \pi \oint _{B_I} d\bar z \, \psi _{\bar z} 
& = & 2 \pi i \int _\Sr \!\!\! d^2u \,  \om_I(u) \, \psi_{\bar u} (u)
\no \\
\int _x ^y dz \int _\Sr \!\!\! d^2u \,  \om (z,u) \, \psi_{\bar u} - 2 \pi \int _x ^y d\bar z \, \psi _{\bar z} 
& = & \int _\Sr \!\!\! d^2u \,  \om _{yx} (u) \, \psi _{\bar u}
\eea
Here, $\om_{yx}(u) $ is the third kind Abelian differential with simple poles in $u$ at $y$ and $x$
with respective residues $+1$ and $-1$ and vanishing $A$-periods (see Appendix A). 
Using these careful definitions of the conditionally convergent integrals needed to evaluate the bosonic periods,  we readily confirm the normalizations of $\hat \om _I$ and $\rh _\eta$ on the bosonic $A$-periods. The bosonic $B$-periods may be evaluated in the same way, and the results will be exhibited in \S \ref{sec3.7} below.

\subsection{Calculation of the fermionic periods of $(\hat\omega_I|\rh_\eta)$}

We begin by recalling the necessary ingredients to compute the fermionic periods
involving the Szeg\"o kernel $S(z,w)$ which was defined in terms of the matrix $M=M_0$ in (\ref{3d11}).
Given that $M_0$ satisfies (\ref{3d9}), formula (\ref{Snorm}) simplifies to give, 
\bea
\label{3f1}
\lim _{z \to p_\eta} (z-p_\eta)^\half  S (z,w) 
+  \sqrt{-1}  \lim _{z \to q_\eta} (z-q_\eta)^\half  S (z,w)
& = & 0
\no \\
\lim _{z \to p_\eta} (z-p_\eta)^\half  S (z,w) 
-  \sqrt{-1}  \lim _{z \to q_\eta} (z-q_\eta)^\half  S (z,w)
& = & - 2 \, \mN \,  h _\eta (w)
\eea
where $h_\eta$ and $\mN$ were defined respectively in (\ref{1h3}) and (\ref{mN}). Using the result on the first line, we readily confirm the normalizations of the fermionic $A$-periods of $\hat \om_I$ and $\rh_\eta$.
Using the result on the second line for the calculation of the fermionic $w^B _\eta$ periods of $\hat \om _I$, we find, 
\bea
w^B _\eta (\hat \om _I) 
=
{ 1 \over \pi}  \int _\Sr \!\!\! d^2w \, h_\eta 
\left ( \half  \chi  \, \hat \om _{Iw} + \mu \, \p_w \hat \om _{I+} + \half (\p_w \mu) \, \hat \om _{I+} \right ) 
\eea
Comparison with the expression for $\hat \Omega _{I\eta}$, we find the following equality,
\bea
w^B _\eta (\hat \om _I) = - { i \over \pi} \oint _{B_I} \rh _\eta
\eea
which is a manifestation of the Riemann bilinear relations for super Riemann surfaces,
as has been explained by Witten. The equality is easy to verify to first order in $\chi$ 
and zero-th order in $\mu$, and may be proven in general by iterated solution of the 
integral equations for the components of $\hat \om _I$ and $\rh_\eta$.
The fermionic $B$-periods $w^B_\eta(\rh_\zeta)$ may be evaluated by the same methods,
and will be exhibited in \S \ref{sec3.7} below.

\subsection{The super period matrix in the supergravity formalism}
\label{sec3.7}

In view of the identification $(\sigma _I | \nu_\eta) \leftrightarrow (\hat \om _I | \rh _\eta)$, the entries of the super period matrix are given by the periods of the latter by
\bea
\label{periods}
\hat\Omega_{IJ} = \oint_{B_J}\hat \omega_I
& \hskip 1in &
\hat\Omega_{I \eta}=w^B _\eta (\hat \om _I) 
\no \\
\hat\Omega_{\eta I} = \oint_{B_I} \rh _\eta
&&
\hat\Omega_{\eta\zeta}=w^B_\zeta  (\rh _\eta)
\eea
As was shown by Witten in \cite{new.Witten}, the super period matrix is symmetric in the $\ZZ_2$-graded sense, so that its components satisfy the relations $\hat \Omega _{JI}=\hat \Omega _{IJ}$, $\hat \Omega _{I\eta} =\hat \Omega _{\eta I}$, and $\hat \Omega _{\eta \zeta} = - \hat \Omega _{\zeta \eta}$. Collecting the results derived above, we find the following expressions for its components, 
\bea
\label{per1}
\hat \Omega _{IJ} &=&
\Omega_{IJ}
 + i  \int _\Sr \!\!\! d^2w \,  \omega_J \left ( \mu \, \hat\omega_{Iw}+
{1\over 2}\chi \,  \hat \omega_{I+} \right )
 \no \\
\hat \Omega _{ \eta I}
 &=&
i  \int _\Sr \!\!\! d^2 w \, \omega_I \left ( \mu \, \rh_{\eta w}
+{1\over 2}\chi \,  \rh_{\eta +} \right )
\no \\
\hat \Omega _{\eta \zeta} & = & (M_0)_{\eta  \zeta} 
+ { 1 \over  \pi}  \int_\Sr \!\!\! d^2 w \, h_\eta 
\left ( \half  \chi  \, \rh_{\zeta w} + \mu \, \p_w \rh_{\zeta +} + \half (\p_w \mu) \, \rh_{\zeta +} \right )
\eea
Recall that the matrix $M_0$ is given explicitly by (\ref{3d9}) in terms of the matrix $\Lambda$, which is itself defined in (\ref{hhprime}), so everything in the above formulas is explicit. 
The validity of the Riemann relations may also be verified on these expressions using the iterated solutions of the integral equations for $\hat \om_I$ and $\rh _\eta$ of (\ref{1h5}) and (\ref{1h4}) respectively.
For example, if we assume that $\mu={\cal O}(\chi^2)$, then the lowest order non-trivial corrections are given as follows,
\bea
\hat \Omega _{IJ}
&=& \Omega _{IJ} + i \int _\Sr \!\!\! d^2w \, \mu \, \om_I \om _J 
-{ i \over 8 \pi} \int _\Sr \!\!\! d^2v  \int _\Sr \!\!\! d^2 w \,
\om _I(v) \chi(v) S (v,w) \chi(w) \om_J (w) + \cO(\chi^4) 
\nonumber\\
\hat \Omega_{\eta I} 
&=&
{ i \over 2} \int _\Sr  \!\!\! d^2 w \, \chi   \om_I  h_\eta + \cO(\chi^3)
\no \\
\hat \Omega _{\eta \zeta} & = & (M_0)_{\eta  \zeta} 
+ { 1 \over  2 \pi}  \int_\Sr \!\!\! d^2 w \, \mu 
\left ( h_\eta  \p_w h_\zeta  - h_\zeta \p_w h_\eta  \right )
\no \\ && 
+ { 1 \over 16 \pi^2}  \int_\Sr \!\!\! d^2 u  \int _\Sr \!\!\! d^2v \,  h_\eta (u) \chi(u) \om (u,v) \chi (v) h)\zeta (v) + \cO(\chi^4)
\eea
The $\ZZ_2$-graded symmetry is now manifest. The expression for the pseudo-classical block $\hat \Omega _{IJ}$  is formally identical to its expression in the absence of R-punctures upon substitution  of the Szeg\"o kernel with R-punctures.

\subsection{Complex structure on $\Sigma_{\rm red}$ and superholomorphicity}

The parametrization of super Riemann surfaces $\Sigma$ used in \S 2 and \S 3 starts 
from the reduced surface $\Sr$ supplemented with a gravitino field $\chi$ and a Beltrami 
differential $\mu$. The period matrix $\Omega_{IJ} $ of $\Sr$ and the pseudoclassical 
block $\hat \Omega_{IJ}$ of the super period matrix do not, in general, agree.
In fact, for genus $g\geq 4$, and arbitrary gravitino field $\chi$, there will, in general,
be no choice of $\mu$ that can achieve $\hat \Omega _{IJ} = \Omega_{IJ}$ since 
the ordinary period matrix $\Omega_{IJ}$ must satisfy the Schottky relations
but, in general, $\hat \Omega_{IJ}$ will not.

\sm

For $g \leq 3$, however, there are no Schottky relations, and for arbitrary $\Omega_{IJ}$
and $\chi$, it will always be possible to find a Beltrami differential $\mu$ such that 
$\hat \Omega_{IJ} = \Omega_{IJ}$. In fact, the condition to be enforced on $\mu$ may be 
read off from (\ref{per1}), and is given by, 
\bea
\label{3k1}
 \int _\Sr \!\!\! d^2 w \, \omega_J \left ( \mu \, \hat\omega_{Iw} +
{1\over 2}\chi \,  \hat \omega_{I+} \right ) =0
\eea
for all $I,J=1, \cdots, g$. In turn, this condition implies that the $(0,1)$ form 
$\mu \, \hat\omega_{Iw} + {1\over 2}\chi \,  \hat \omega_{I+}$ must be 
in the range of the $\p_{\bar w}$ operator acting on functions, so that we have,
\bea
\mu \, \hat\omega_{Iw} +
{1\over 2}\chi \,  \hat \omega_{I+} = - \p_{\bar w} \lambda _I
\eea
for some $g$-component  function $\lambda _I $.
The functions $\lambda _I$ may be reconstructed explicitly from this equation, 
up to an additive constant, and we have,
\bea
\lambda _I (u) - \lambda _I (v) = { 1 \over 2 \pi} \int _\Sr \!\!\! d^2w \, 
\om _{uv} \left ( \mu \, \hat\omega_{Iw} +
{1\over 2}\chi \,  \hat \omega_{I+} \right ) 
\eea
The addition to $\om_{uv}$ of a linear
combination of holomorphic forms $\om_I$ is immaterial  in view of (\ref{3k1}).
This property guarantees that $\lambda_I(u)$ are well-defined and single-valued on $\Sr$. 

\sm

The function $\lambda_I$ provides a powerful connection relating the superholomorphic
structure given by the 1/2 forms $\hat \omega_I$ of the super Riemann surface $\Sigma$
to the complex structure given by the 1-forms $\om_I$ of the reduced surface $\Sr$.
To see this, we first establish the following relation, by inspection of (\ref{1h5}), 
\bea
\label{omhatom}
\hat \om _{Iz}  (z) = \om_I(z) + \p_z \lambda _I (z)
\eea
The correspondence may be expressed directly in superspace language, and we have,
\bea
\hat \om _I (z, \theta) & = & \theta \, \om _I (z) + \cD_+ \Lambda _I (z,\theta)
\no \\
\Lambda_I  (z, \theta) & = & \lambda _I (z) + \theta \, \hat \om _{I+}  (z)
\eea
It is now immediate from these expressions that the period matrices $\hat \Omega _{IJ}$
and $\Omega _{IJ}$ coincide, since the differentials on which they are built differ just by
a total derivative. Relations of the form (\ref{omhatom}) were instrumental in the study of NS scattering amplitudes \cite{D'Hoker:2005}. They can be expected to play a similar role in the study of Ramond scattering amplitudes.

\subsection{Additional Periods}
\label{sec3.9}

In addition to the periods of super holomorphic 1/2 forms reviewed earlier in this section, 
it is natural to consider further periods obtained as line integrals of superholomorphic 
1/2 forms between pairs of R-punctures.\footnote{We are grateful to Edward Witten for sharing his unpublished note \cite{Witten.note} on additional periods with us.} In particular, such periods emerge automatically upon the non-separating degeneration of a super Riemann surface when the degenerating 
cycle carries a spin structure associated with a Ramond state, as will be shown in \S \ref{sec5}. 
The additional periods may be either even or odd, and  their description will depend upon the picture chosen 
to describe Ramond states. In picture -1/2,  R-punctures are characterized by a 
Ramond divisor whose  fiber is of dimension $ 0|1$ at each puncture. Supersymmetry
transformations act by automorphism on the fiber but leave the divisor invariant.

\sm

The additional periods may have practical applications to parametrize the moduli 
space of super Riemann surfaces with a small number of R-punctures. The reason is
that the superperiod matrix $(\hat \Omega _{IJ}, \hat \Omega _{I\eta}, \hat \Omega_{\eta \zeta})$
does not, for low $r$, have a sufficient number of independent parameters to account for
all the even and odd moduli of $\mM _{g,0,2r}$. For example, the number of odd entries 
$\hat \Omega _{I\eta}$ equals $gr$ while the number of odd moduli in picture -1/2 is $2g-2+r$.
For $g \geq 2$, the super period marix provides enough odd moduli only when $r \geq 2$.
Similarly, the entries of $\hat \Omega _{\eta \zeta}$ provide enough even moduli to account
for the $2r$ R-punctures only when $r \geq 5$. Thus, for surfaces of low $r$, it may be convenient
to have additional periods available.

\sm

We shall now provide a brief account of the additional periods for general genus $g$ but 
restrict to the simplest case $r=1$ which is perhaps the most important one.
We denote the two R-punctures by $p,q$. To facilitate the description of the additional periods, 
it will be convenient to specify not just the divisors $p, q$, but also auxiliary points $\theta_p$ 
and $\theta _q$  in each one of the $0|1$ fibers. In the -1/2 picture, the variables 
$\theta_p, \theta_q$ transform under supersymmetry by automorphism of the fiber, and 
should therefore not be thought of as moduli.

\sm

The additional even periods $\hat V_I$  and odd period $\hat V_0$ are given as 
follows,
\bea
\label{3m1}
\hat V_I = \int _{q, \theta_q} ^{p, \theta _p} \hat \om _I
\hskip 1in 
\hat V_0 = \int _{q, \theta_q} ^{p, \theta _p} {\hat \rho_0 \over \sqrt{2} \mN} 
\eea
They may be evaluated in components using (\ref{1h5}) and (\ref{1h4}) giving the even periods,
\bea
\label{3m2}
\hat V_I = \int ^p _q \om _I + {1 \over 2 \pi} \int _\Sr \!\!\! d^2w \, \om _{pq}  \left (
\mu \, \hat \om _{Iw} + \half \chi \, \hat \om _{I+} \right ) 
+ \theta _p w^p (\hat \om_I) - \theta _q \, w^q (\hat \om _I)
\eea
and the odd period,
\bea
\label{3m3}
\sqrt{2} \, \mN \, \hat V_0 =  {1 \over 2   \pi} \int _\Sr \!\!\! d^2w \, \om _{pq}  \left (
\mu \, \rh _{0w} + \half \chi \, \rh _{0+} \right ) 
+ \theta _p w^p (\rh_0) - \theta _q \, w^q (\rh _0)
\eea
Here, $\om _{pq}(z) $ is the third kind Abelian differential with vanishing $A$-periods, as reviewed in (\ref{A9}).
It is manifest that the additional periods $\hat V_I$ and $\hat V_0$ have non-trivial monodromy as $p$ and $q$ are taken around non-trivial homology cycles on the surface, and we have,
\bea
\label{3m5}
p \to p +A_J & \hskip 0.7in & \hat V_I \to \hat V_I + \delta _{IJ}, \hskip 0.52in \hat V_0 \to \hat V_0
\no \\
p \to p +B_J &  & \hat V_I \to \hat V_I + \hat \Omega _{IJ}, \hskip 0.5in \hat V_0 \to \hat V_0 + \hat \Omega _{0I}
\eea
The dependence on the fermionic periods in (\ref{3m2}) and (\ref{3m3}) may be rendered explicit
using the results of (\ref{wpq}) and (\ref{wpq0}). The additional even periods become,
\bea
\label{1j6}
\hat V_I = \int ^p _q \om _I 
+ {1 \over 2 \pi} \int _\Sr \!\!\! d^2w \, \om _{pq} \left (  \mu \, \hat \om _{Iw} + \half \chi \, \hat \om_{I+} \right )
+  \theta _B  \hat \Omega _{I0} 
\eea
while  the additional odd period becomes, 
\bea
\label{1j7}
\hat V_0 =  {1 \over 2 \sqrt{2} \pi \mN } \int  _\Sr \!\!\! d^2 w \, \om_{pq}  \left (  \mu \, \rh _{0w} + \half \chi \, \rh_{0+} \right ) 
+ \theta _A 
\eea
Here, we have found it natural to introduce the notations, 
\bea
\label{1j8}
\theta _A & = & { 1 \over \sqrt{2}}  ( \theta _p  + \sqrt{-1} \, \theta _q )
\no \\
\theta _B & = &  { 1 \over \sqrt{2}}  ( \theta _p  - \sqrt{-1} \, \theta _q )
\eea
Under the supersymmetry transformations of (\ref{susy}) on $\chi$ and $\mu$, 
the additional periods $\hat V_I, \hat V_0$ are invariant provided the points $\theta _p, \theta_q$
in the $0|1$-dimensional fibers transform as follows,
\bea
\theta _p & \to & \theta _p + \xi (p)
\no \\
\theta _q & \to & \theta _q + \xi (q)
\eea
We note that the additional periods are invariant under diffeomorphisms that leave
the R-punctures invariant.

\sm

In genus $2$ with two R-punctures, it is useful to consider also deformations $\mu$ which fix at the same time the additional even periods $\hat V_I$ introduced in (\ref{3m1}).

\newpage

\section{Variations of Periods}
\setcounter{equation}{0}
\label{sec4}

In the absence of  R-punctures the genus $g=2$ super period matrix  provides a holomorphic 
projection of the moduli space $\mM_{2,0,0}$ of genus 2 super Riemann surfaces onto the 
moduli space $\cM_2$ of ordinary Riemann surfaces, and  has provided a valuable tool
for the construction of the superstring measure and scattering amplitudes.
In the presence of two R-punctures, Witten has shown \cite{new.Witten} that the genus 2
super period matrix provides a natural meromorphic projection of supermoduli space $\mM_{2,0,2}$ 
onto $\cM_2$, which may be used to derive powerful results on the fate of the vacuum energy 
in general heterotic superstring compactifications which enjoy $\cN=1$ tree-level supersymmetry. 
Some of these results appear to generalize to genus 3 as well \cite{new.Witten}.

\sm

There is good reason to expect that the super period matrix with R-punctures will play a 
valuable role in the evaluation of a measure and scattering amplitudes with R-punctures 
at genus 2 as well. For low values of $r$, additional periods may be required, as explained in \S \ref{sec3.9}. Once these periods 
have been chosen to provide a parametrization by coordinates collectively denoted $m^A$
of the moduli space $\mM_{2,0,2r}$  of super Riemann surfaces of genus 2 with $2r$ punctures, 
then a key ingredient in the construction  of the measure and scattering amplitudes is the super 
Beltrami differential  $H_A=\bar \theta \p ( - \mu  - \theta \chi )/ \p m^A$ associated with the choice of 
coordinates $m^A$, and the  space of superholomorphic 3/2 differentials $\Phi_B$ dual to $H_A$. 
In this section we shall evaluate  $H_A$ and $\Phi_B$. As applications may exist to 
genus 2 and genus 3 in superstring theory for an arbitrary number of R-punctures, we shall
calculate the variational formulas of Witten's super period matrix for general $g$ and $r$,
and supplement this with the variational formulas for the additional periods for the case $r=1$.

\subsection{Summary of variational formulas}

The derivations of the variational formulas are rather technical, so we shall begin by giving an overview of the results, and relay derivations to later subsections. We shall consider the  variation  generated by deformations $\delta \chi$ of $\chi$ and $\delta \mu$ of $\mu$ of the various entries of the super period matrix. For the case $r=1$ we shall also evaluate the variations of the additional periods. The most succinct  formulation of the variational formulas is in superspace language.  Introducing the superspace Beltrami differential,
\bea
\label{4a1}
H = \bar \theta ( - \delta \mu - \theta \delta \chi)
\eea
the deformations of the components of the super period matrix are found to be  as follows,
\bea
\label{4a2}
\delta \hat \Omega _{IJ} = \oint _{B_J} \delta \hat \om _I 
& = & \< H | \Phi_{IJ} \>
\no \\
\delta \hat \Omega _{\eta I}  = \, \oint _{B_I} \delta \rh _\eta 
& = & \< H | \Phi _{\eta I} \>
\no \\
\delta \hat \Omega _{\eta \zeta }  = w^B _\zeta (\delta \rh _\eta )
& = & \< H | \Phi _{\eta \zeta } \>
\eea
where the super holomorphic 3/2 forms are given by,
\bea
\label{4a3}
\Phi _{IJ} & = & { i \over 2} \hat \om _I (\cD_+ \hat \om _J) + { i \over 2} (\cD_+ \hat \om _I) \hat \om _J
\no \\
\Phi _{\eta I} & = & { i \over 2} \rh _\eta (\cD_+ \hat \om _I) - { i \over 2} (\cD_+ \rh _\eta) \hat \om _I
\no \\
\Phi _{\eta \zeta} & = & { i \over 2} \rh _\eta (\cD_+ \rh_\zeta) - { i \over 2} (\cD_+ \rh _\eta) \rh _\zeta
\eea
Their superholomorphicity, namely $\cD_- \Phi _{IJ}= \cD_- \Phi _{\eta I}= \cD_- \Phi _{\eta \zeta}=0$,
follows from the superholomorphicity of $\hat \om _I$ and $\rh _\eta$.
The structure of these variations  in terms of superholomorphic 3/2 forms is analogous to the structure found without R-punctures in \cite{D'Hoker:2001zp} for genus 2.

\sm

The variations of the additional periods may be computed as well. Here, we shall do so only
for the  case  $r=1$, which is perhaps of greatest physical relevance. 
One obtains, 
\bea
\label{4a4}
\delta \hat V_I & = & { 1 \over 2 \pi i} \< H | \Psi _I^{pq} \> +  \delta \left ( \theta _B \hat \Omega _{I0} \right )
\no \\
\delta \hat V_0 & = & { 1 \over 2 \sqrt{2}  \pi i \mN } \< H | \Psi _0^{pq} \> + \delta \theta _A
\eea
where $\Psi ^{pq} _I$ and $\Psi^{pq}_0$ are now super meromorphic 3/2 forms which have poles 
at $p$ and $q$. They may be expressed in terms of super holomorphic and meromorphic
1/2 forms, 
\bea
\label{4a5}
\Psi^{pq}_I & = & { i \over 2} \hat \om ^{pq} (\cD_+ \hat \om _I) + { i \over 2} (\cD_+ \hat \om ^{pq}) \hat \om _I
\no \\
\Psi^{pq}_0 & = & { i \over 2} \hat \om ^{pq}  (\cD_+ \rh_0) + { i \over 2} (\cD_+ \hat \om ^{pq} ) \rh _0
\eea
where $\hat \om ^{pq}$ is the superholomorphic 1/2 form with simple poles at $p$ and $q$ with residues 
$+1$ and $-1$, and vanishing $A$-periods. Its explicit form is given in Appendix \ref{B}. Note that we have allowed for variations of $\theta _A$ and $\theta _B$, and thus of the points  in the Ramond divisors.

\subsection{Varying superholomorphic 1/2 forms}

Deformations of $\chi $ and $\mu$ produce variations $\delta \hat \om _I$ and $\delta \rh _\eta$ 
of the superholomorphic 1/2 forms  $\hat \om _I$ and $\rh _\eta$. To compute these variations
in terms of existing data, it is convenient to start from the normalization conditions, and differential 
equations which these variations must satisfy. The normalization conditions express the fact that
we require the normalizations of the bosonic and fermionic $A$-periods to remain unchanged,
\bea
\label{4b1}
\int _{A_J} \delta \hat \om _I = \oint _{A_I} \delta \rh _\eta =0
 \hskip 1in   
w^A_\eta (\delta \hat \om _I)=w^A_\eta ( \delta \rh _\zeta ) = 0
\eea
The components of the variation $\delta \hat \om= \delta \hat \om _+ + \theta \delta \hat \om _z$ 
of any one of the superholomorphic 1/2 forms $\hat \om= \hat \om _+ + \theta \hat \om _z$
satisfy a set of differential equations derived by variation of  (\ref{1c2}), 
\bea
\label{4b2}
\left ( \pbz + \mu \, \pz + \half (\pz \mu) \right ) \delta \hat \om _+ + \half \chi \, \delta \hat \om _z 
& = & \tau_{z \bar z} ^+
\no \\
\pbz \delta \hat \om_z + \p_z \left ( \half \chi \, \delta \hat \om_+ + \mu \, \delta \hat \om _z \right ) 
& = &  - \p_z \tau _{\bar z}
\eea 
where the sources are given by,
\bea
\label{4b3}
\tau_{z \bar z} ^+ & = & - \delta \mu \, \pz \hat \om _+ - \half (\pz \delta \mu) \hat \om _+ 
- \half \delta \chi \,  \hat \om _z
\no \\
\tau _{\bar z} & = & 
 \half \delta \chi \,  \hat \om_+ + \delta \mu \, \hat \om _z
\eea
This system of differential equations may be solved for the deformations $ \delta \hat \om _+ $ and  $ \delta \hat \om _z$ in terms of three kernels $\cG_0(z,w), \cG_+(z,w)$, and $\cG_1(z,w)$ which are meromorphic sections of the bundles $K_z\otimes K_w$, $\cR^{-1}_z \otimes K_w$, and $\cR^{-1}_z \otimes \cR^{-1}_w$ respectively, and will be more completely defined and calculated below and  in Appendix \ref{B}. The integral equation corresponding to (\ref{4b2})  is given by,
\bea
\label{2b6}
\delta \hat \om _+ (z) & = & \delta \varpi _+  (z) 
+ { 1 \over 2 \pi} \int _\Sr \!\!\! d^2w \,  \Big ( \cG_1(z,w) \,\tau_{w \bar w} ^+(w) + \cG_+ (z,w) \, \tau _{\bar w}  \Big )
\no \\
\delta \hat \om _z (z) & = & \delta \varpi _z  (z) 
+ { 1 \over 2 \pi} \int _\Sr \!\!\! d^2w \,  \Big ( \cG_+(w,z) \, \tau_{w \bar w} ^+ (w) + \cG_0 (z,w) \, \tau _{\bar w}  \Big )
\eea
Expressing the combinations $\tau_{w \bar w} ^+$ and $\tau _{\bar w} $ in terms of $\delta \chi$
and $\delta \mu$ provides us with the desired variational formulas.
The forms $\delta \varpi _+  (z) $ and $\delta \varpi _z  (z)$
are the components of an arbitrary super holomorphic 1/2 form, to be fixed by normalization.
In Appendix \ref{B}, a suitable set of normalization conditions will be imposed on the kernels $\cG_0, \cG_+$, and $\cG_1$, in such a manner that one can set $\delta \varpi _z=\delta \varpi _+=0$.

\subsection{Varying the super period matrix}

Combining these results yields the general formulas for the deformations of $B$-periods,
\bea
\oint_{B_I} \delta \hat \om & = & 
i \int _\Sr \!\!\! d^2w \, \Big ( \hat \om _{I+} \tau_{w \bar w} ^+ + \hat \om _{Iw} \tau _{\bar w} \Big )
\no \\
w^B_\eta (\delta \hat \om) & = &
{ 1 \over 2 \pi} \int _\Sr \!\!\! d^2w \, \Big ( \rh _{\eta w} \tau_{\bar w} - \rh _{\eta +} \tau_{w \bar w} ^+ \Big)
\eea
Expressing $\tau$ in terms of the variations $\delta \chi $ and $\delta \mu$, and recasting the result in 
superspace with the help of (\ref{4a1}) and the superspace expressions for super holomorphic 1/2 forms,
\bea
\hat \om (z, \theta) & = & \hat \om _{+}(z) + \theta \, \hat \om _{z} (z)
\no \\
\hat \om _I(z, \theta) & = & \hat \om _{I+}(z) + \theta \, \hat \om _{Iz} (z)
\no \\
\rh _\eta (z, \theta) & = & \rh _{\eta +}(z) + \theta \, \rh _{\eta z} (z)
\eea
we see that the combinations
\bea
\hat \om _I (\cD_+ \hat \om) + (\cD_+ \hat \om _I ) \, \hat \om
& = & 
\hat \om _{I+} \hat \om _z + \hat \om _{Iz} \hat \om_+ 
+ \theta \Big ( 2 \hat \om _{Iz} \hat \om_z + \p_z \hat \om _{I+} \hat \om_+ 
- \hat \om _{I+} \p_z \hat \om _+ \Big )
\no \\
\rh _\eta (\cD_+ \hat \om ) - (\cD_+ \rh _\eta ) \, \hat \om 
& = & 
\rh _{\eta +} \hat \om _z - \rh _{\eta z} \hat \om_+ 
+ \theta \Big ( 2 \rh _{\eta z} \hat \om_z - \p_z \rh _{\eta +} \hat \om_+ 
+ \rh _{\eta +} \p_z \hat \om _+ \Big )
\quad
\eea
will yield the superholomorphic 3/2 forms listed in the results of (\ref{4a2}) and (\ref{4a3}), upon setting
$\hat \om$ equal to either $\hat \om_I$ or $\rh _\eta$.

\subsection{Varying additional periods}

We shall now prove the variational formulas for the additional periods in the case $r=1$. They are given by the line integral (\ref{integ}) between points $\theta_p$ and $\theta _q$ in the Ramond divisors at $p,q$ applied to the 1/2 forms $\hat \om_I$ and $\rh_0$.  These periods take the form,
\bea
\label{2j1}
\hat V_I = \int ^{p,\theta_p} _{q,\theta_q} \hat \om_I  
\hskip 1in 
\hat V_0 = {1 \over \sqrt{2} \,  \mN} \int ^{p,\theta_p} _{q,\theta_q} \rh_0  
\eea
The deformations of $\hat V_I$ and $\hat V_0$ receive contributions from the variation of $ \hat \om$ due to the deformations of $\chi$ and $\mu$, as well as from the variation of the points $\theta_p, \theta_q$ in the Ramond divisor, and we have,
\bea
\label{4d1}
\delta \hat V_I & = &  
 \int ^p _q \left (  dz \,  \delta \hat \om _{Iz} - \half d\bar z \, \chi \, 
\delta \hat \om _{I+} - d \bar z \, \mu \, \delta \hat \om _{Iz} \right ) + \delta \left ( \theta_B \hat \Omega _{I0} \right )
\no \\
\delta \hat V_0 & = &  
{1 \over \sqrt{2} \mN}  \int ^p _q \left (  dz \,  \delta \rh _{0z} - \half d\bar z \, \chi \, 
\delta \rh _{0+} - d \bar z \, \mu \, \delta \rh _{0z} \right ) + \delta  \theta_A 
\eea
Making use of the integrals of the kernels in (\ref{A12}), we find, 
\bea
\delta \hat V_I = 
{1 \over 2 \pi}  \int _\Sr \!\!\! d^2w \, 
\Big ( \hat \om _+^{pq} \tau_{w \bar w} ^+ + \hat \om _w^{pq} \tau _{\bar w} \Big )
+ \delta \left ( \theta_B \hat \Omega _{I0} \right )
\eea
where the form $\hat \om$ in $\tau$ equals $\hat \om _I$, and 
\bea
\delta \hat V_0 =  
{1 \over 2 \pi}  \int _\Sr \!\!\! d^2w \, 
\Big ( \hat \om _+^{pq} \tau_{w \bar w} ^+ + \hat \om _w^{pq} \tau _{\bar w} \Big )
 + \delta  \theta_A 
\eea
where the form $\hat \om$ in $\tau$ now equals $\rh_0/\sqrt{2}\mN$. Expressed in superspace with the help of $\hat \om ^{pq}(z,\theta) = \hat \om ^{pq}_+ (z) + \theta \hat \om ^{pq}_z (z)$, these results produce
(\ref{4a4}) and (\ref{4a5}).

\newpage

\section{The Super Period Matrix from Degenerations}
\setcounter{equation}{0}
\label{sec5}

In this section, we examine the limit of the super period matrix of a super Riemann surface of genus $g+1$, in the non-separating limit with Ramond monodromy around the degenerating cycle. The limiting surface is then a super Riemann surface of genus $g$ with $2$ R-punctures. Our goal is to show that the limit of the super period matrix in genus $g+1$ consists indeed of the bosonic components of the limiting surface of genus $g$ with $2$ R-punctures. The case of a limiting surface with $2r$ R-punctures for $r>1$ can then be considered as a degeneration of a surface with $2(r-1)$ R-punctures. A simultaneous degeneration of $r$ cycles in a surface with no punctures would not be generic.

\subsection{The super period matrix in genus $g+1$}

We consider the following one-parameter family of super Riemann surface $(\mg,\chi)$ in genus $g+1$ with no puncture. Let $\Sigma$ be a Riemann surface of genus $g$, and construct the family of degenerating surfaces $\Sigma^\ep$ of genus $g+1$, as $\ep\to 0$, by the classical construction described in Fay \cite{Fay} for ordinary Riemann surfaces. That is, we choose neighborhoods $U_p$ and $U_q$ around two points $p$ and $q$, which can be identified with the unit disk through coordinates $z_p$ and $z_q$ respectively. The family of  surfaces $\Sigma^\ep$ is constructed by
\bea
\label{7h1}
\Sigma^\ep=\Sigma\setminus (\ep U_p\cup\ep U_q)/\sim
\eea
where the equivalence relation $\sim$ means that we identify the annuli 
$\{|\ep|<|z_p|<1\}$ and $\{|\ep|<|z_q|<1\}$ around $p$ and $q$ through the relation
\bea
\label{7h2}
z_p z_q=\ep.
\eea
Note that for any fixed $\rho$ with $0<\rho<1$,  and all $0<\ep<\rho$, the surfaces $\Sigma^\ep$ all contain the open set $\Sigma \setminus (\rho U_p\cup\rho U_q)$. 

\sm

For super Riemann surfaces $\Sigma$, it was shown in section 6 of \cite{Witten:2012ga} that, for Ramond degenerations,  the identification of the even coordinates in (\ref{7h2}) needs to be supplemented with an identification of the entire Ramond divisors $\cF_p$ and $\cF_q$ at $p$ and $q$. In terms of coordinates $\theta _p$ and $\theta_q$ for these divisors, the identification will specify the combination $\theta _q  \pm \sqrt{-1} \, \theta _p$,
where the two sign choices produce the two possible relations between the spin structure of $\Sigma ^\ep$ and the generalized spin structure of $\Sigma$ (see section 6.2 of  \cite{Witten:2012ga}).

\sm

We fix an even spin structure $\delta$ on $\Sigma^\ep$ and consider a non-separating degeneration
of type R to a surface $\Sigma$ with R-punctures $p,q$. Let $\chi$ be a gravitino field on $\Sigma$, and let $\chi^\ep$ be gravitino fields on $\Sigma^\ep$ which converge to $\chi$ on compact subsets of $\Sigma\setminus \{p,q\}$, and which remain uniformly bounded in a neighborhood of $p$ and $q$. We fix a canonical homology basis $(A_{\cal I},B_{\cal I})$ for $\Sigma^\ep$ with $B_{g+1}$ the non-separating cycle. Let $\omega_{\cal I}^\ep$ be the corresponding basis of holomorphic forms dual to $A_{\cal I}$, and let $\Omega_{\cal IJ}^\ep$ the corresponding period matrix
\bea
\label{7b0}
\oint_{A_{\cal I}}\omega_{\cal J}^\ep=\delta_{\cal IJ},
\hskip 1in 
\oint_{B_{\cal I}}\omega_{\cal J}^\ep=\Omega_{\cal IJ}^\ep.
\eea
The super period matrix $\hat\Omega_{\cal I J}^\ep$, with $\cI, \cJ =1, \cdots, g+1$ was derived in \cite{D'Hoker:1989ai}. Let $\hat\omega_{\cal I}^\ep = \hat\omega_{{\cal I} +}^\ep + \theta\hat\omega_{{\cal I} z}^\ep$ be the superholomorphic 1/2 forms for $(\mg,\chi)$ given by the integral equations,\footnote{for simplicity, we shall set $\mu=0$ in this section; the results in the presence of $\mu$ are analogous.}
\bea
\label{g+1}
\hat\omega_{{\cal I}+}^\ep
&=&-{1\over 4 \pi}
\int_{\Sigma^\ep} d^2w\, S^\ep[\delta](z,w) \chi^\ep(w) \, \hat\omega_{Iw}^\ep(w)
\nonumber\\
\hat\omega_{{\cal I}z}^\ep
&=&\omega_{\cal I}^\ep
+{1\over 4\pi}\int_{\Sigma^\ep}d^2w\, \omega^\ep(z,w) \chi^\ep(w) \, \hat\omega_{I+}^\ep(w)
\eea
where $\omega^\ep(z,w)=\p_z\p_w\ln E^\ep(z,w)$ and $S^\ep[\delta](z,w)$ are respectively the prime form and the Szeg\"o kernel for the surface $\Sigma^\ep$. Then the super period matrix is given by
\bea
\label{g+1a}
\hat\Omega_{\cal IJ}^\ep
=
\Omega_{\cal IJ}^\ep
+ {i \over 2} 
\int_{\Sigma^\ep}d^2w\, \omega_{\cal I}^\ep(w) \, \chi^\ep (w)  \hat\omega_{{\cal J}+}^\ep(w).
\eea
Equations (\ref{g+1}) for $\hat\omega_{\cal I}$ can be solved iteratively by expanding in powers of the nilpotent field $\chi$. It suffices then to compute the degeneration limits of $\omega_{\cal I}^\ep$, $\Omega_{\cal IJ}^\ep$, $E^\ep(z,w)$, and $S^\ep[\delta](z,w)$.

\subsection{The degeneration limits of $\omega_{\cal I}^\ep$,
$\Omega_{\cal IJ}^\ep$, and $E^\ep(z,w)$}

These degeneration limits are well-known, and we shall just quote the results from \cite{Fay}.
Let $z\in \Sigma\setminus (U_p \cup U_q)$. Then to order ${\cal O}(\ep)$,
\bea
\label{7b1}
\om_I^\ep (z) & = & \om _I (z) + \cO (\ep) 
\no \\
\om _\mh^\ep  (z) & = & { 1 \over 2 \pi i} \om _{qp}(z) + \cO (\ep).
\eea 
Here $\omega_I$ are the holomorphic differentials of the surface $\Sigma$, and $\om_{qp}(z)$ is the meromorphic differential with simple poles at $p$ and $q$ on $\Sigma$. Whenever convenient, we shall use the shorthand $\mh=g+1$. The period matrix on $\Sigma ^\ep$ is then given by,
\bea
\label{7b2}
\Omega ^\ep _{\cI \cJ} = 
\left ( \matrix{ \Omega _{IJ} & a_I \cr a_J & (\ln \ep)/2 \pi i  + c \cr } \right )+ \cO (\ep)
\hskip 1in a_I = - \int ^p _q \om_I
\eea
where $\Omega_{IJ}$ is the period matrix of $\Sigma$ and $c$ depends only on $\Omega$,
and not on $p,q$. Similar formulas hold for the prime form, and its derivatives, for $z,w\in \Sigma\setminus (U_p \cup U_q)$,
\bea
\label{7b4}
\ln E^\ep (z,w) & = & \ln E(z,w) + \cO (\ep),
\no\\
\p_z \ln E^\ep (z,w) & = & \p_z \ln E(z,w) +\cO( \ep) 
\no \\
\p_z \p_w \ln E^\ep (z,w) & = & \p_z \p_w \ln E(z,w) +\cO(\ep).
\eea

\subsection{The degeneration limit of $S^\ep[\delta](z,w)$}

Evaluating the non-separating degeneration limit of the Szeg\"o kernel is the main step in calculating the limit of the super period matrix. We decompose each even spin structures $\delta$ on the surface $\Sigma ^\ep$ 
in accord with the non-separating degeneration along the cycle $B_\mh=B_{g+1}$, 
\bea
\label{spindeg}
\delta = \left [ \matrix{ \kappa ' & \kappa '' \cr \delta'_\mh & \delta '' _\mh \cr} \right ]
\hskip 1in 
\kappa = [\kappa ' ~ \kappa '']
\eea
Here, $\kappa$ parametrizes the generalized spin structure induced by $\delta$ on $\Sigma$, and 
$\delta _\mh', \delta _\mh ''  \in \{ 0, \half \}$. The Szeg\"o kernel on $\Sigma ^\ep$ is given by,
\bea
\label{7c3}
S^\ep [\delta] (z,w) = { \tet [\delta ] (z-w, \Omega ^\ep) 
\over \tet [\delta ] (0, \Omega ^\ep) E^\ep (z,w)}
\eea
on the surface $\Sigma ^\ep$.  We already know that $E^\ep (z,w) = E(z,w) + \cO(\ep)$. The degenerations of the  $\tet$-functions are obtained as follows.

\subsubsection{Degeneration of $\tet [\delta] (0,\Omega^\ep)$}

We compute the degeneration of $\tet [\delta ] (0, \Omega^\ep) $ using the series representation 
of (\ref{B1}) with the sum running over $n_\cI$ with $\cI=1, \cdots ,  g+1=\mh$. Since 
$\Omega ^\ep _{\mh \mh} \to + i \infty$, the leading  contributions in the series arise as follows. 
When $\delta '_\mh =0$ the degeneration is of the Neveu-Schwarz type, only the term $n_\mh=0$ contributes,  and the $\tet$-constant converges to, 
\bea
\tet [\delta ] (0, \Omega_\ep) = \tet [\kappa ] (0, \Omega) + \cO(\ep)
\eea
The characteristic $\kappa$ speficies the spin structure induced by $\delta$ on the surface $\Sigma$.
When $\delta _\mh '=\half$, the degeneration is of the Ramond type, both the terms $n_\mh =0,-1$ contribute, and the $\tet$-constant converges to,
\bea
\label{7c4}
\tet [\delta ] (0, \Omega_\ep) =
2 e^{i \pi \delta _\mh ''} e^{i \pi \Omega _{\mh \mh} /4} \tet [\kappa] \left ( \half p-\half q , \Omega \right )
+ \cO(\ep)
\eea
The characteristic $\kappa$ together with the positions of the points $p$ and $q$ in the $2^{2g}$-fold cover of $\Sigma$, specifies the generalized spin structure induced by $\delta$ on $\Sigma$. Formula (\ref{7c4}) is valid as long as this leading contribution is non-vanishing, namely $\kappa +(p-q)/2 \not \in \Theta$, where $\Theta$ is the $\tet$-divisor. In terms of $\ep$, this leading term is of order $\ep ^{1 \over 8}$ and corrections 
to it will be suppressed by integer powers of $\ep$.

\subsubsection{Degeneration of $\tet [\delta] (z-w,\Omega^\ep)$}

We compute the degeneration of $\tet [\delta ] (z-w, \Omega^\ep) $ using the series representation 
of (\ref{B1}) as well.  To leading order, the Abelian integrals involving $z,w$ are given by, 
\bea
\label{7c5}
v^\ep_I & = & \int _w^z \om ^\ep _I = \int _w^z \om _I + \cO(\ep)
\no \\
u^\ep & = & \int _w ^z \om ^\ep _\mh = { 1 \over 2 \pi i} \ln { E(z,q) E(w,p) \over E(z,p) E(w,q)} + \cO(\ep)
= {i \over \pi} \ln \phi _{pq} (z,w)+ \cO(\ep)
\eea
where the holomorphic covering function $\phi$ of (\ref{7c5}) is given by, 
\bea
\phi _{pq} (z,w) = \left ( { E(z , p) E(w,q) \over E(z, q) E(w, p)}  \right )^\half
\eea
Decomposing the summation in the series for the $\tet$-function in accord with the degeneration 
along the cycle $B_\mh$, we have,
\bea
\tet [\delta](z-w, \Omega ^\ep) & = &
\sum _{m \in \bZ} \, \, \sum _{n -\kappa '  \in \bZ^g} \exp \Big \{ 
i \pi n^t \Omega n + i \pi \Omega ^\ep _{\mh \mh} (m + \delta '_\mh)^2
\\ &&  
+ 2 \pi i n^t (v^\ep + a (m+\delta '_\mh) + \kappa '')
 + 2 \pi i (m + \delta _\mh ') ( u^\ep + \delta _\mh '') \Big \}
\no
\eea
As the limits when $\ep \to 0$ of the quantities $u^\ep, v^\ep_I$ are finite, but
$\Omega _{\mh \mh} \to  + i \infty$, the leading contribution to the limit will
result from the following terms in the summation over $m$. When $\delta _\mh'=0$, it is just
$m=0$ which contributes, and we obtain,
\bea
\tet [\delta](z-w, \Omega ^\ep) = \tet [\kappa ] (z-w, \Omega) + \cO(\ep)
\eea
When $\delta '_\mh = \half$, then $m=0,-1$ contribute, and to leading order, we find,
\bea
\label{7c6}
\tet [\delta](z-w, \Omega ^\ep) & = &
e^{i \pi \Omega _{\mh \mh} + i \pi  \delta '' _\mh} \, \phi _{pq}(w,z) \,
\tet [\kappa] \left ( z - w - \half p + \half q ; \Omega \right )
\no \\ &&
+ e^{i \pi \Omega _{\mh \mh} - i \pi  \delta '' _\mh} \, \phi _{pq} (z,w) \, 
\tet [\kappa] \left ( z - w + \half p- \half q ; \Omega \right )
\eea
where we have used $\exp ( i \pi u^\ep )= \phi _{pq} (w,z)$ in view of (\ref{7c5}).

\subsubsection{Degeneration of the Szeg\"o kernel $S^\ep [\delta](z,w;\Omega ^\ep)$}

Assembling all ingredients, we find that for NS non-separating degenerations with $\delta _\mh'=0$, the 
Szeg\"o  kernel with even spin structure $\delta$ on $\Sigma ^\ep$ simply tends to the Szeg\"o kernel with even spin structure $\kappa$ on $\Sigma$,  
\bea
\label{7c23}
S^\ep [\delta ] (z,w;\Omega ^\ep)  = S [\kappa ] (z,w; \Omega) + \cO (\ep)
\eea
For Ramond degenerations with $\delta _\mh'=\half$, the Sezg\"o kernel with even spin structure $\delta$ on $\Sigma ^\ep$ tends to the following limit,
\bea
S^\ep [\delta ] (z,w;\Omega ^\ep)  =
\half S^{pq} (z,w) - \half S^{pq}(w,z)+\cO(\ep).
\eea
This relation is obtained by assembling the limit of the Szeg\"o kernel in (\ref{7c3}) from the limits of its factors  
in (\ref{7c6}), (\ref{7c4}), and (\ref{7b4}, and identification with the Szeg\"o kernel $S^{pq}(z,w)$ given in (\ref{1e6}). With the help of the definition (\ref{1e8}) this result in turn is seen to coincide with the anti-symmetric Szeg\"o kernel in the presence of two Ramond punctures $p,q$,
\bea
S^\ep [\delta ] (z,w;\Omega ^\ep) = S(z,w) +\cO(\ep).
\eea

\subsection{The limit of the super period matrix}

We shall now determine the limit of $\hat\Omega_{IJ}^\ep$, with $I,J=1,\cdots, g+1=\mh$, as $\ep\to 0$. 
We begin by recalling some basic relations between the moduli spaces $\mM_{g,n,2r}$ of super Riemann surfaces of genus $g$, with $n$ Neveu-Schwarz punctures, and $2r$ Ramond punctures. For the case of interest, namely $g \geq 2$, their dimensions are $3g-3+n+2r|2g-2+n+r$.

\sm

In the supergravity formulation of super Riemann surfaces, the $2g$ odd moduli of  $\mM_{g+1,0,0}$  are encoded in the field $\chi^\ep$. As a result, the expansion in powers of $\chi^\ep$ of the even quantities such as the differentials $\hat \om^\ep _{\cI z}$ in (\ref{g+1}) and the super period matrix $\hat \Omega ^\ep_{\cI \cJ}$ of (\ref{g+1a}) vanish beyond $\cO((\chi^\ep) ^{2g})$, while the odd quantities such as  $\hat \om ^\ep_{\cI +}$ in (\ref{g+1}) vanish beyond $\cO((\chi ^\ep)^{2g-1})$. 

\sm

Upon a non-separating degeneration of type NS or  R  of a surface in $\mM_{g+1,0,0}$,  one even modulus becomes the degeneration parameter $\ep$, while  two even moduli become the bosonic coordinates of the punctures. For the NS degeneration, two odd moduli of $\chi^\ep$ become the odd coordinates of the NS punctures. 

\sm

For the R degeneration, the fate of the odd moduli is more delicate, as the odd dimension of $\mM_{g+1,0,0}$ is one higher than the odd dimension of its naive limit $\mM_{g,0,2}$. This subtlety was  carefully explained and resolved in section 6.2 of \cite{Witten:2012ga}. The Ramond degeneration divisor $\mD$ equals not $\mM_{g,0,2}$ but rather a $\CC^{0|1}$ line bundle over $\mM_{g,0,2}$ with additive structure group $\CC^{0|1}$. In the supergravity formulation, this means that the $2g$ odd moduli in $\mM_{g+1,0,0}$ which span $\chi^\ep$ will redistribute into the $2g-1$ odd moduli of $\mM_{g,0,2}$ which span $\chi$  along with one odd coordinate $\alpha$ which parametrizes the $\CC^{0|1}$ fiber  of the bundle $\cD \to \mM_{g,0,2}$. The parameter $\alpha $ enters into the prescription for gluing the Ramond divisors $\cF_p$ and $\cF_q$ with coordinates $\theta_p$ and $\theta_q$ respectively through the relation, 
\bea
\theta _p \pm \sqrt{-1} \, \theta _q =\alpha
\eea
In the supergravity formulation of super Riemann surfaces, $\chi$ will depend on $\alpha$, and 
$\alpha$ will transform under supersymmetry by a shift.

\sm

We now return to the degeneration limits of the super period matrix for $\mM_{g+1,0,0}$ and even spin structure. For non-separating degenerations of type NS, the limit found in (\ref{7c23}) for the Szeg\"o kernel on $\Sigma^\ep$ gives the Szeg\"o kernel on $\Sigma$, and thus reduces the components $\hat \Omega _{I J} ^\ep$ for $I, J = 1, \cdots, g$ of the super period matrix on $\Sigma ^\ep$ with even spin structure $\delta$ to the components $\hat \Omega _{I J}$ of the super period matrix for the corresponding surface $\Sigma $ with even spin structure $\kappa$, related by (\ref{spindeg}). The components $\hat \Omega ^\ep _{I \mh}$ reduce to  line integrals between the two NS punctures, and $\hat \Omega _{\mh \mh}$ diverges as $\ln \ep$,  as was the case for bosonic surfaces in (\ref{7b2}).

\sm

For non-separating degenerations of type R, we focus attention first on the $g \times g$ block of the super period matrix with components $\hat \Omega ^\ep _{IJ}$ with $I,J=1,\cdots, g$. All the ingredients that enter into the expressions for these components in (\ref{g+1a}), namely the holomorphic forms $\omega_I^\ep(z)$, gravitino field $\chi^\ep(z)$, Szeg\"o kernel $S^\ep[\delta](z,w)$, and prime form $E^\ep(z,w)$ converge to their respective counterparts $\omega_I(z)$,  $\chi(z)$,  $S(z,w)$, and  $E(z,w)$ on the super Riemann surface $\Sigma$ with R-punctures $p$ and $q$ when $z,w$ are outside the funnel  region $\{|\ep|<|z_p|<\rho\}$. As a result, the components of the bock $\hat \Omega ^\ep _{IJ}$ with $I,J=1,\cdots, g$ converge to the pseudo-classical block with components $\hat \Omega_{IJ}$,
of the surface $\Sigma$ with R-punctures $p,q$,
\bea
\label{5m2}
\lim _{\ep \to 0} \hat\Omega_{IJ}^\ep 
=
\hat\Omega_{IJ} \hskip 1in I,J =1 \cdots, g
\eea
Since  $\hat \Omega_{IJ}$ is invariant under any supersymmetry transformation, including those
which do not vanish at $p$ and $q$, it is clear that any dependence in $\chi$ upon the gluing parameter $\alpha$ will cancel out in the limit $\hat \Omega _{IJ}$, which thus effectively only depends upon $2g-1$ odd moduli.

\sm

For non-separating degenerations of type R, the limit of $\hat \Omega _{\mh I}^\ep$ is obtained from (\ref{g+1a}) using the limit (\ref{7b1}). Naively, this would lead to the following expression for $\hat \Omega _{\mh I}$,  
\bea
\label{5m1}
- \int ^p _q \om _I - { 1 \over 4 \pi} \int _\Sr \!\!\! d^2w \, \om _{pq}(w) 
\chi (w) \, \hat \om _{I+}(w)
\eea 
This result is naive because the funnel will now contribute due to the presence of the poles at the punctures $p,q$ caused by the differential $\om _{pq}$. Considering the behavior under local supersymmetries, we see that (\ref{5m1}) is invariant under all supersymmetries which vanishes at $p,q$, but not under supersymmetries which do not vanish at $p,q$. For the latter, the contribution of the funnel is non-zero, but we shall not attempt here to evaluate it. Since we have earlier identified the line integrals $\hat V_I$ as additional periods which are invariant under all supersymmetries, it is compelling to conjecture that the correct limit, taking into account the contributions from the funnel, will give, 
\bea
\lim _{\ep\to 0}
\hat\Omega_{I\mh}^\ep
=
\hat V_I
\eea
where $\hat V_I$ are the additional bosonic periods defined by the line integrals from $p$ to $q$ in (\ref{3m1}). Also, the component $\hat \Omega _{\mh \mh} ^\ep$ diverges like $\ln \ep$ as was the case for bosonic surfaces in (\ref{7b2}). 

\sm

Finally, note that the limits (\ref{5m2}) and (\ref{5m1}) demonstrate that the full super period matrix at genus $g$ with two R-punctures may be obtained from the non-separating degeneration of the super period matrix at genus $g+1$ without R-punctures. This is manifest for the quasi-classical block $\hat \Omega _{IJ}$ of (\ref{5m2}), while the odd components $\Omega _{0I} $ may be obtained from (\ref{5m1}) by taking the derivative with respect to $\theta _B$ in the explicit expression for $\hat V_I$ obtained in (\ref{1j6}), where $\theta_B$ parametrizes the single odd modulus of the surface of genus $g+1$ which has become pure gauge upon the non-separating degeneration.

\newpage

\appendix

\section{$\tet$-functions, Abelian differentials, and prime form}
\setcounter{equation}{0}
\label{A}

In this appendix, we review Jacobi $\tet$-functions, the prime form, and Abelian differentials
and their basic properties on a  compact Riemann surface genus $g$ (see \cite{Fay} and for example \cite{Verlinde:1986kw,D'Hoker:1988ta} for applications to string perturbation theory).

\subsection{$\tet$-functions}

The $\tet$-function is an entire function in a $g \times g$ symmetric matrix $\Omega \in \CC^{g \times g}$
whose imaginary part is positive definite, a column matrix $\zeta \in \CC^g$, and characteristics $\kappa = [\kappa ' \, \kappa '']$ with column matrices $\kappa ' , \kappa '' \in \CC^g$, defined by
\be
\label{B1}
\tet [\kappa ] (\zeta, \Omega)
\equiv
\sum _{n \in {\bf Z}^g} \exp \Big \{\pi i (n+\kappa ')^t \Omega (n+\kappa')
+ 2\pi i (n+\kappa ')^t (\zeta + \kappa '') \Big \}
\ee
For arbitrary characteristics $\kappa$ and arbitrary $\zeta$, there is a redundancy in the definition of the $\tet$-function, which results in the following relation valid for arbitrary  $\lambda \in \CC^{2g}$, 
\bea
\label{B2}
\tet [\kappa + \lambda ] (\zeta , \Omega )
=
\tet [\kappa ](\zeta + \lambda '' + \Omega \lambda' , \Omega) \ \exp \Big \{ i
\pi (\lambda ')^t  \Omega \lambda ' + 2 \pi i (\lambda ')^t (\zeta +\lambda '' +
\kappa '')  \Big \}
\eea
The following monodromy relations hold for any $M', M'' \in {\bf Z}^g$, as special cases of (\ref{B2}), 
\bea
\label{B3}
\tet [\kappa ] (\zeta + M'' + \Omega M', \Omega )
&=&
\tet [\kappa ](\zeta , \Omega) \ \exp \{ -i \pi (M')^t \Omega M'
- 2 \pi i (M'')^t (\zeta +\kappa '' - \kappa ')  \}
\nonumber \\
\tet [\kappa ' +M', \kappa '' +M'' ] (\zeta , \Omega)
&=&
\tet [\kappa] (\zeta, \Omega) \ \exp \{ 2\pi i (\kappa ')^t  M'' \}
\eea
Of interest here is $\kappa$ associated with a spin structure or, in the presence of R-punctures a generalized spin structure, in which case $\kappa$ are  half-integer characteristics, so
that $\kappa ' , \kappa '' \in \left \{ 0, \half \right \} ^g$. Here, $\tet$ is even or odd in $\zeta$ depending on whether  $4 (\kappa ')^t \kappa ''$ is even or odd.

\subsection{Function theory on a compact Riemann surface}

The basic objects of function theory on a compact Riemann surface $\Sigma_0$ (we shall reserve the customary notation $\Sigma$ for super Riemann surfaces) of genus $g$ are the Abelian differentials and the prime form. To define them, we choose a canonical homology basis in $H^1(\Sigma_0, \ZZ)$ given by cycles $A_I, B_I$ with canonical intersection form $\# (A_I, A_J)=\#(B_I, B_J) =0$ and $\#(A_I, B_J) = \delta _{IJ}$ for $I,J=1,\cdots, g$. Modular transformations on the basis $A_I, B_I$ are defined to leave the intersections invariant, and form the group $Sp(2g, \ZZ)$.

\sm

The holomorphic Abelian differentials $\om _I$ for $I=1,\cdots, g$ are holomorphic sections of the canonical buncle $K$ on $\Sigma_0$ which are normalized on $A$-cycles, and whose integrals on $B$-cycles produce the period matrix $\Omega _{IJ} $ of the surface $\Sigma_0$,
\bea
\label{B4}
\oint _{A_J} \om _I = \delta _{IJ} 
\hskip 1in 
\oint _{B_J} \om _I = \Omega _{IJ} 
\eea
The Jacobian is defined by $J(\Sigma_0) = \CC^g / ( \ZZ^g + \Omega \ZZ^g)$. Given a base point $z_0$ on $\Sigma$, the Abel map is defined for $d$ points $z_i$ and  multiplicities $n_i$ with $i=1,\cdots, d$ by 
\bea
\label{B5}
n_1 z_1 + \cdots + n_d z_d = \sum _{i=1}^d n_i \int _{z_0} ^{z_i} \om _I
\eea
For integer multiplicities $n_i $, the Abel map is single-valued on $J(\Sigma_0)$.  

\sm

To define the bundle of spinors $K^\half$ requires specifying a spin structure. On a surface of genus $g$ there are $2^{2g}$ spin structures, of which $2^{g-1}(2^g+1)$ are even, while $2^{g-1}(2^g-1)$ are odd, depending on whether the number of holomorphic sections of $K^\half$ is even or odd. Having fixed a reference spin structure, all other spin structures may be labelled by a half-characteristic $\kappa$ which is even or odd in accord with the parity of the spin structure. Generically, there is one holomorphic section $h_\kappa (z)$ of $K^\half $ for $\kappa $ odd, given by the formula, 
\bea
\label{B6}
h_\kappa (z)^2 = \sum _{I=1}^g \om _I(z) \p_I \tet [\kappa ] (0, \Omega)
\eea
up to an overall sign. The holomorphic 1-form on the right side admits a single-valued square root in view of the fact that all of its zeros are double. For $\kappa$ even, it clearly vanishes.

\sm

The prime form $E(z,w)$ is a $(-1/2,0)$-form in each variable $z$ or $w$, defined on the universal cover of the surface $\Sigma_0$ in terms of an arbitrary odd characteristic $\kappa$, by
\bea
\label{B7}
E(z,w) = { \tet [\kappa ] (z-w,\Omega) \over h_\kappa (z) h_\kappa (w)}
\eea
$E(z,w)$ is odd under the interchange of $z$ and $w$, and satisfies $E(z,w)=z-w+{\cal O}(z-w)^3$. Its monodromy around $A$-cycles is trivial, while around $B$-cycles its monodromy is given by
\bea
\label{B8}
E(z+B_I,w) = E(z,w)\, {\rm exp} \left ( -\pi i\Omega_{II}+2\pi i\int_z^w\omega_I \right ).
\eea
The Abelian differentials $\om (z,w) $ of the second kind, and $\om _{xy}(z)$ of the third kind, with unit residues and vanishing $A$-periods, may be expressed in terms of the prime form by,
\bea
\label{B9}
\om (z,w)= \p_z \p_w \ln E (z,w) \hskip 1in \om _{xy}(z) = \p_z \ln { E(z,x) \over E(z,y)}
\eea
Their $B$-periods are given as follows,
\bea
\label{B10}
\oint _{B_I} dz \, \om (z,w) = 2 \pi i \om _I(w) 
\hskip 1in 
\oint _{B_I} dz \, \om_{xy} (z) = \int ^x _y \om _I
\eea
Abelian differentials of second and third kind obeying other normalizations may be obtained by the addition of Abelian differential $\om_I$ of the first kind to $\om(z,w)$ and $\om _{xy}(z)$ respectively.

\subsection{Deformation theory on a compact surface}

We shall need the following simple formulation of the deformation of a complex structure by a Beltrami differential $\mu$ on a purely bosonic Riemann surface $\Sigma _0$ (such as for example the reduced surface $\Sr$): a complex structure is specified by the class of its holomorphic functions. If the holomorphic functions $f(z)$ for a given complex structure are specified by $\p_{\bar z}f=0$, and if 
$\mu= \mu_{\bar z}{}^z$ is a Beltrami differential, then the holomorphic functions of the deformed complex structure are defined by the equation
\bea
\label{deformedbard}
(\p_{\bar z}+\mu\p_z)f=0.
\eea
This means that the deformation is implemented by the basic shift for derivatives
\bea
\label{deformedbard1}
\p_{\bar z} \to \p_{\bar z}+\mu\p_z
\eea
on scalars, with similar shifts on tensors of arbitrary weights.  Corrections of the form $\mu \bar \mu$ may be ignored since all conformal anomalies cancel in critical string theory, while corrections involving higher powers of $\mu$ alone are absent. An alternative formulation is by introducing coordinates $w$ which are holomorphic with respect to the new complex structure, that is, which satisfy the equation (\ref{deformedbard}). The change $z\to w$ is a quasi-conformal transformation, and function theory can then be formulated in terms of $w$ and $\p_{\bar w}$. 

\sm

In concrete calculations of scattering amplitudes it will turn out to be much more convenient to keep the coordinates $z$ on the super Riemann surface $\Sigma$ fixed and associated with the complex structure of $\Sr$.

\sm

The variation $\delta _{ww} \phi$ of any object $\phi$ under the variation of the complex structure of the surface $\Sigma _0$ by a Beltrami differential $\mu$ is defined as follows,
\bea
\delta \phi = { 1 \over 2 \pi} \int _{\Sigma _0} d^2 w \, \mu _{\bar w} {}^w \delta _{ww} \phi
\hskip 0.8in \mu _{\bar w} {}^w = \half g_{w \bar w} \delta g^{ww}
\eea
The variations of the basic functional objects on the surface $\Sigma _0$ that will be needed here are given as follows,
\bea
\delta _{ww} \Omega _{IJ} & = & 2 \pi i \om_I (w) \om _J(w)
\no \\
\delta _{ww} \om _I(x)  & = & \om _I (w) \p_x \p_w \ln E(x,w)
\no \\
\delta _{ww} \ln E(x,y) & = & - \half \Big (  \p_w \ln  E(w,x) - \p_w  E(w,y) \Big )^2
\no \\
\delta _{ww} S(x,y) & = & \half \p_w S(x,w) S(w,y) - \half S(x,w) \p_w S(w,y)
\eea
On the last line, $S(x,y)$ stands for the Szeg\"o kernel on a surface $\Sigma _0$ without R-punctures.

\newpage

\section{Calculation of Variational Formulas}
\setcounter{equation}{0}
\label{B}

In this appendix we shall provide some  details of the variational calculations in \S \ref{sec4}, and spell out the equations governing the kernels $\cG_0 (z,w)$, $\cG_+ (z,w) $, and $\cG_1 (z,w) $ which are meromorphic sections of the bundles $K_z\otimes K_w$, $\cR^{-1}_z \otimes K_w$, and $\cR^{-1}_z \otimes \cR^{-1}_w$ respectively.

\subsection{Differential and integral equations for the kernels}

By inspection of the defining equations (\ref{2b6}) and (\ref{4b2}), the  kernels must satisfy the following differential equations, 
\bea
\label{A4}
\left ( \pbz + \mu (z) \pz + \half (\pz \mu)(z)  \right ) \cG_1(z,w)  + \half \chi (z) \cG_+(w,z)  
& = & 2 \pi \delta (z,w)
\no \\
\pbz \cG_+(w,z)  + \p_z \left ( \half \chi (z) \cG_1(z,w)  + \mu (z) \cG_+(w,z) \right ) 
& = &  0
\eea 
as well as,
\bea
\label{A5}
\left ( \pbz + \mu (z) \pz + \half (\pz \mu)(z) \right ) \cG_+(z,w)  + \half \chi (z) \cG_0(z,w)  
& = & 0
\no \\
\pbz \cG_0 (z,w)  + \p_z \left ( \half \chi (z) \cG_+(z,w)  + \mu (z) \cG_0(z,w) \right ) 
& = &  - 2 \pi \p_z \delta (z,w)
\eea 
The solutions to these differential equations are, of course, not unique, as one is free to add linear combinations of the super holomorphic 1/2 forms. One particular solution to equations (\ref{A4}) and (\ref{A5}) is given by,
\bea
\label{A6}
\cG_0 \, (z,w) & = &
\om (z,w) + { 1 \over 2 \pi} \int _\Sr \!\!\! d^2u \,  \om (z,u) \left ( \mu(u) \cG_0(u,w) 
+ \half \chi (u) \cG_+ (u,w) \right )
\\
\cG_+(z,w) & = & - { 1 \over 2 \pi} \int _\Sr \!\!\! d^2 u \,  S(z,u) \left ( \half \chi (u) \cG_0(u,w) 
+ \left ( \mu(u) \p_u + \half (\p_u \mu)(u) \right ) \cG_+(u,w) \right )
\no \\
\cG_+(w,z) & = & 
{ 1 \over 2 \pi} \int _\Sr \!\!\! d^2 u \,  \om (z,u) \left ( \mu(u) \cG_+(w,u) 
+ \half \chi (u) \cG_1 (u,w) \right )
\no \\
\cG_1 \, (z,w) & = & S(z,w) - { 1 \over 2 \pi} \int _\Sr \!\!\! d^2 u \,  S(z,u) \left ( \half \chi (u) \cG_+(w,u) 
+ \left ( \mu(u) \p_u + \half (\p_u \mu)(u) \right ) \cG_1(u,w) \right )
\no
\eea
One may view this solution to  (\ref{A4}) and (\ref{A5}) as resulting from a specific choice of normalizations, by fixing the bosonic and fermionic $A$-periods. This route will be adopted below.

\sm

We note that the kernels may be viewed as the components of the double derivative $\cD_+^\bz \cD_+ ^\bw \ln \cE(\bz, \bw)$ of the super prime form $\cE(\bz,\bw)$, generalized to the case with R-punctures from its definition in the case without R-punctures in \cite{D'Hoker:1989ai}.

\subsection{Normalization of $A$-periods}

For the bosonic $A$-periods, the solution of (\ref{A6}) leads to, 
\bea
\label{A7}
\oint _{A_I} du \, \cG_0 (u,w) - 2 \pi \oint _{A_I} d\bar u \left ( \mu(u) \cG_0(u,w) 
+ \half \chi (u) \cG_+ (u,w) \right )
& = & 0
\no \\
\oint _{A_I} du \, \cG_+ (w,u) - 2 \pi \oint _{A_I} d\bar u \left ( \mu(u) \cG_+(w,u) 
+ \half \chi (u) \cG_1 (u,w) \right )
& = & 0
\eea
while for the fermionic $A$-periods, it leads to, 
\bea
\label{A8}
\lim _{z \to p_\eta} (z-p_\eta)^\half \cG_+(z,w) + \sqrt{-1} \lim _{z \to q_\eta} (z-q_\eta)^\half \cG_+(z,w) & = & 0
\no \\
\lim _{z \to p_\eta} (z-p_\eta)^\half \cG_1(z,w) + \sqrt{-1} \lim _{z \to q_\eta} (z-q_\eta)^\half \cG_1(z,w) & = & 0
\eea
The combination of these conditions specifies the kernels uniquely. As a result, the bosonic and fermionic $A$-periods of the variation $\delta \hat \om$ of an arbitrary superholomorphic 1/2 form in (\ref{2b6}) will automatically obey the normalization conditions of (\ref{4b1}), as was our goal.

\subsection{Calculation of $B$-periods}

Next, we evaluate the $B$-periods of the kernels. Using the $B$-period of $\om(z,w)$ in (\ref{B10}), we find the following expressions for the bosonic $B$-periods of the kernels $\cG$,
\bea
\label{A9}
\oint _{B_I} du \, \cG_0 (u,w) - 2 \pi \oint _{B_I} d\bar u \left ( \mu(u) \cG_0(u,w) 
+ \half \chi (u) \cG_+ (u,w) \right )
& = &  2 \pi i \hat \om_{Iw}(w) 
\no \\
\oint _{B_I} du \, \cG_+ (w,u) - 2 \pi \oint _{B_I} d\bar u \left ( \mu(u) \cG_+(w,u) 
+ \half \chi (u) \cG_1 (u,w) \right )
& = & 
2 \pi i \hat \om _{I+}(w)
\quad
\eea
Similarly, using the fermionic $B$-period of $S(z,w)$ of (\ref{3f1}) we find the following
expressions for the fermionic $B$-periods of the kernels,
\bea
\label{A10}
\lim _{z \to p_\eta} (z-p_\eta)^\half \cG_+(z,w) - \sqrt{-1} \lim _{z \to q_\eta} (z-q_\eta)^\half \cG_+(z,w) & = & 
+ \rh_{\eta w}(w)
\no \\
\lim _{z \to p_\eta} (z-p_\eta)^\half \cG_1(z,w) - \sqrt{-1} \lim _{z \to q_\eta} (z-q_\eta)^\half \cG_1(z,w) & = & 
- \rh _{\eta+} (w)
\eea
The proof of the formulas for the fermionic $B$-periods is straightforward. For the 
integral $B$-periods, one first proves the following expressions,
\bea
\label{A11}
&&
\oint _{B_I} dz \, \cG_0(z,w) - 2 \pi \oint _{B_I} d\bar u \left ( \mu(u) \cG_0(u,w) 
+ \half \chi (u) \cG_+ (u,w) \right )
\no \\ 
&& \hskip 0.5in =
2 \pi i \om _I(w) + i \int _\Sr \!\!\! d^2u \,  \om _I(u) \left ( \mu(u) \cG_0(u,w) 
+ \half \chi (u) \cG_+ (u,w) \right )
\no \\ &&
\oint _{B_I} dz \,  \cG_+(w,z) - 2 \pi \oint _{B_I} d\bar u \left ( \mu(u) \cG_+(w,u) 
+ \half \chi (u) \cG_1 (u,w) \right ) 
\no \\ && \hskip 0.5in  = 
i \int _\Sr \!\!\! d^2u\,  \om _I(u)  \left ( \mu(u) \cG_+(w,u) 
+ \half \chi (u) \cG_1 (u,w) \right )
\eea
By comparing these expressions with those for $\hat \om _I$ given in (\ref{1h5}), both expanded in powers of $\chi$ and $\mu$,  it is immediate to establish identifications of (\ref{A9}), as illustrated below in \S \ref{secB4}.

\sm

Finally, there are the line integrals between two R-punctures, which we shall consider here only for $r=1$. 
The line integrals are given as follows, 
\bea
\label{A12}
\int_q ^p du \, \cG_0 (u,w) - 2 \pi \int _q ^p d\bar u \left ( \mu(u) \cG_0(u,w) 
+ \half \chi (u) \cG_+ (u,w) \right )
& = &   \hat \om^{pq} _{w}(w) 
\no \\
\int_q ^p du \, \cG_+ (w,u) - 2 \pi \int_q ^p d\bar u \left ( \mu(u) \cG_+(w,u) 
+ \half \chi (u) \cG_1 (u,w) \right )
& = & 
 \hat \om ^{pq}_+(w)
\quad
\eea
One shows that the components of the super meromorphic 1/2 form $\hat \om ^{pq}(z,\theta)=\hat \om _+^{pq}(z) + \theta \hat \om ^{pq}_z(z)$ satisfy the following differential equations, 
\bea
\label{2j14}
\left ( \pbw + \mu  \pw + \half (\pw \mu) \right ) \hat \om ^{pq}_+(w)  
+ \half \chi \hat  \om ^{pq}_w (w)   & = & 0
\no \\
\pbw \hat \om^{pq}_w (w)  + \p_w \left ( \half \chi  \hat  \om^{pq}_+(w) + \mu  \hat \om ^{pq}_w (w) \right ) 
& = &  2 \pi \Big ( \delta (w,r)-\delta (w,s) \Big )
\eea 
Therefore, $\hat \om ^{pq}$ is a super meromorphic 1/2 form with simple poles at $p$ and $q$
with residues $+1$ and $-1$ respectively, and no fermionic residues.
We note that its periods are given by, 
\bea
\oint _{A_I} \hat \om ^{pq}=0 
\hskip 1in 
\oint _{B_I} \hat \om ^{pq}= 2 \pi i \hat V_I
\eea
The fermionic periods may be deduced as well, and we have,
\bea
w^A (\hat \om ^{pq}) = - \theta _B 
\hskip 1in
w^B (\hat \om ^{pq}) = \hat V_0 - \theta_A
\eea

\subsection{Proofs}
\label{secB4}

Full proofs may be given by an iterative solution of the integral equations for the kernel and for the differentials. Considering for simplicity the special case where $\mu=0$, a proof may be formulated by developing a full matrix-operator expansion of the quantities at hand. The starting point is to recast the integral equations of (\ref{A6}) in a matrix operator form, using the  shorthand $\hat \chi = \chi/4\pi$,  and we have,
\bea
\cG_+ = - S \hat \chi \cG_0
& \hskip 1in &
\cG_0 = \om + \om \hat \chi \cG_+
\no \\
\cG_+^t  = + \om \hat \chi \cG_1
& \hskip 1in &
\cG_1 = S - S \hat \chi \cG_+^t
\eea
Pairwise elimination allows us to solve these equations schematically as follows,
\bea
\cG_+ = - ( 1 + S \hat \chi \om \hat \chi ) ^{-1} \, S \hat \chi \om
& \hskip 1in &
\cG_0 = ( 1 + \om \hat \chi S \hat \chi ) ^{-1} \, \om
\no \\
\cG_+^t = + ( 1 + \om \hat \chi S \hat \chi ) ^{-1} \, \om \hat \chi S
&&
\cG_1 =( 1 + S \hat \chi \om \hat \chi ) ^{-1} \, S
\eea
We use the basic transposition rules $ \om ^t = \om $ and $ S^t = - S$.
To show that the two expressions on the left are consistent with one another, it suffices to
expand both in powers of $\hat \chi$, 
to use the transposition rules $(S\hat \chi \om \hat \chi)^t = \hat \chi \om \hat \chi S$
and $(\om \hat \chi S \hat \chi)^t = \hat \chi S \hat \chi \om$, and to rearrange the results.
It is manifest that these two lines are transposes of one another.

\end{document}